\RequirePackage[l2tabu]{nag} 
\documentclass[10pt]{article}

\usepackage[utf8]{inputenc}
\usepackage[T1]{fontenc}
\usepackage[top=1.3in, bottom=1.4in, left=1.4in, right=1.4in]{geometry}

\usepackage{lmodern}
\usepackage[tracking=true,kerning=true,final]{microtype}
\usepackage{bbm}

\usepackage{bm}
\usepackage{mathtools}
\usepackage{amssymb}

\usepackage[dvipsnames,]{xcolor}
\usepackage[unicode=true,pdfusetitle, bookmarks=true,bookmarksnumbered=false,bookmarksopen=false, breaklinks=false,hidelinks,colorlinks=true]{hyperref}
\hypersetup{colorlinks=true, linkcolor=MidnightBlue, citecolor=MidnightBlue, urlcolor=MidnightBlue}

\usepackage{caption}   
\usepackage{subcaption}
\usepackage{color}
\usepackage{graphicx}
\graphicspath{{./figures/}}

\usepackage[noabbrev]{cleveref}

\newcommand{\overbar}[1]{\mkern 1mu\overline{\mkern-1mu#1\mkern-1mu}\mkern 1mu}
\DeclarePairedDelimiter\abs{\lvert}{\rvert}

\DeclareMathOperator{\expec}{\mathbb E}

\DeclareMathOperator{\prob}{\mathbb P}
\DeclareMathOperator{\pdf}{\rho}
\DeclareMathOperator{\real}{\text{Re}}

\newcommand{\blank}{\,\cdot\,}
\DeclareMathOperator{\sech}{sech}
\renewcommand{\d}[2]{\frac{d #1}{d #2}} 
\newcommand{\dd}[2]{\frac{d^2 #1}{d #2^2}} 
\DeclareMathOperator{\spn}{span}

\usepackage{amstext}
\usepackage{amsmath}

\begin{document}
\title{A probabilistic decomposition-synthesis method for the quantification of rare events due to internal instabilities}

\author{Mustafa A. Mohamad, Will Cousins, Themistoklis P. Sapsis\medskip\thanks{Corresponding author: \href{mailto:sapsis@mit.edu}{sapsis@mit.edu},
Tel: (617) 324-7508, Fax: (617) 253-8689%
}\\
Department of Mechanical Engineering, 
\\ Massachusetts Institute of Technology, \\
77 Massachusetts Ave., Cambridge, MA 02139}
\date{\today}
\maketitle

\begin{abstract}
We consider the problem of probabilistic  quantification of dynamical systems that have heavy-tailed characteristics. These heavy-tailed features are associated with rare transient responses due to the occurrence of internal instabilities.   Systems  with these properties can be found in a variety of areas including mechanics, fluids, and waves. Here we develop a computational method, a probabilistic decomposition-synthesis technique,  that takes into account the nature of internal instabilities to inexpensively determine the non-Gaussian probability density function   for any arbitrary quantity of interest. Our approach relies on the decomposition of the statistics into a `non-extreme core', typically Gaussian, and a heavy-tailed component. This decomposition is in full correspondence with a partition of the phase space into a `stable' region where we have no internal instabilities, and a region where non-linear instabilities  lead to rare transitions with high probability. We quantify the statistics in the stable region using a Gaussian approximation approach, while the non-Gaussian distributions associated with the intermittently unstable regions of the  phase space are inexpensively computed through order-reduction methods that take into account the strongly nonlinear character of the dynamics. The probabilistic information in the two domains is analytically synthesized through a total probability argument. The proposed approach allows for the accurate quantification of   non-Gaussian tails at  more than $10$ standard deviations, at a fraction of the cost associated with the direct  Monte-Carlo simulations. We demonstrate the probabilistic decomposition-synthesis method for rare events for two dynamical systems exhibiting extreme events: a two-degree-of-freedom system of nonlinearly coupled oscillators, and in a nonlinear envelope equation characterizing the propagation of unidirectional water waves.
\end{abstract}


\clearpage


\section{Introduction}\label{sec:introduction}
Quantifying extreme or rare events is a central issue for many technological processes and natural phenomena. As extreme events, we consider  rare transient responses that  push  the system away from its statistical steady state, which often lead to catastrophic consequences. Complex systems exhibiting rare events include (i) dynamical systems found in nature, such as the occurrence of rare climate events~\cite{hense06, Frei2001, Easterling2000} and  turbulence~\cite{Majda_filter, majda1997, qi15, tong15, Majda2014a}, formation of freak water waves in the ocean \cite{Dysthe08, Olagnon2005, Kharif2003, trulsen1996, cousinsSapsis2015_JFM, cousins_sapsis}; but also (ii) dynamical systems in engineering applications involving mechanical components subjected to stochastic loads~\cite{naess_book, extreme_value09, paik03, Shadman}, ship rolling and capsizing~\cite{belenky07, Kreuzer, arnold_l}, critical events in power grids~\cite{Kishore12, Pourbeik, Susuki2012a, kundur94}, as well as chemical reactions and conformation changes in molecules~\cite{Wigner1938, Pratt1986, Bolhuis2003, E2010}.

For many systems of practical interest like those above, it has now been well established that rare transitions occur frequently enough   that they are of critical importance. These intermittent events are randomly triggered while the system evolves on the (stochastic) background attractor, and they are subsequently governed, primarily, by the spatiotemporally local and strongly nonlinear dynamics associated with finite-time instabilities. Systems with these properties  pose a significant  challenge for uncertainty quantification schemes~\cite{majda_branicki_DCDS} and in recent years a wide range of research efforts has taken place in various fields towards the quantification and short-term prediction of rare events in complex dynamical systems. 

The quantification of rare events is one of the most fundamental problems in chemistry. Chemical reactions, conformation changes of molecules, and quantum tunneling are examples of rare events~\cite{Wigner1938, Pratt1986, Bolhuis2003, E2010}. These events are rare because the system has to overcome certain barriers of energetic or entropic nature in order to move from one stable state to the other. The usual setup for modeling such systems is their formulation in terms of a Langevin equation, i.e. a dynamical system with some non-quadratic potential that has multiple equilibria excited by white noise (see for example~\cite{E2010}). The goal then is to study barrier-crossing events by computing transition rates as well as shortest paths between states. For such systems the classical transition state-theory (TST)~\cite{Eyring1935,  Wigner1938} has been successful in providing the foundation for the development of computational tools that determine transition trajectories between different states. However, important limitations for transition state theory may occur when the system potential is not smooth. In this case it is essential to seek for transition tubes (i.e. ensembles of transition trajectories). The statistical framework to analyze such transition-path ensemble is known as the transition-path theory (TPT)~\cite{E.2006, Metzner2006, Metzner2009} and it has been applied successfully for applied to interesting and challenging problems in a variety of areas, for example, in chemistry, biology, and material science.

Although successful on quantifying transitions between different states, path theory can have limitations when   considering dynamical systems that exhibit rare responses due to the occurrence of intermittent instabilities (as opposed to multiple equilibria) that lead to strong energy transfers between modes, as  is the case in turbulence or nonlinear waves.  In such cases, the rare event is not the result of a transition that takes the system from one state to the other. Rather, it is the result of intermittent instabilities that can `push' the system away from its statistical steady state to a dynamical regime with a strongly transient character. The period that the system spends away from its statistical steady state attractor as well as the distance from the attractor usually takes on a continuous range of values, depending on the intensity and duration of the instability. This situation is completely different from the setup involving transitions between discrete states, for which TST and TPT theories have been developed.

Large deviations theory~\cite{varadhan84, varadhan08, dembo00, stroock84}  is a powerful method for the probabilistic quantification of extreme events in sequences of probability distributions. It has also been applied in the context of stochastic differential equations, known as Freidlin-Wentzell theory~\cite{freidlin98}, as well as for stochastic partial differential equations~\cite{chow92, sowers92, sritharan06}. In this case, the method essentially provides us with rates of convergence to probabilistic limits. For example, for a dynamical system excited by very low intensity noise, large deviations theory   gives closed form expressions bounding the probability for a big deviation of the stochastic solution from the completely unperturbed solution, i.e. a probabilistic characterization of the stochastic solution relative to the asymptotic (deterministic) limit. Despite its importance, it is not straightforward to apply this framework in order to quantify extreme events that rise out of the steady state attractor of the system due to the occurrence of intermittent instabilities, which is the problem that we are interested here.

For the probabilistic description of rare events in phenomena characterized by intermittent instabilities, the analysis is usually limited to the statistical examination of  observed statistics. For example, in ocean engineering, where it is important to analyze the probability of upcrossings and maxima for various quantities of interests (e.g. wave elevation or mechanical stresses), the standard setup involves the adoption of globally stable dynamics, for which many techniques have been developed  (see e.g.~\cite{Naess1982, Soong_Grigoriou93, naess_book}). There are numerous technical steps involved in this case that lead to elegant and useful results, but the starting point is usually the assumption of stationarity in the system response, which is not a valid hypothesis for intermittently unstable systems. 

Extreme value theory~\cite{extreme_value09, Leadbetter, Thomas_extremes, Embrechts12, Galambos} is also a widely applicable method which focuses on thoroughly analyzing the extreme properties of stationary stochastic processes following various distributions. However, even in this case the analysis does not take into account any information about the unstable character of the system dynamics and is usually restricted to very specific forms of correlation functions for the response statistics~\cite{extreme_value09, Leadbetter}. To this end, it is not surprising that   for a large range of complex systems exhibiting intermittent characteristics, the Monte-Carlo method is  the only reliable computational approach to arrive at   accurate  estimates for the tails. However, for high-dimensional systems this direct approach  is usually prohibitively expensive for practical purposes.  

Recently, there have been  efforts to quantify the heavy tail statistical structures of systems undergoing  transient instabilities. In \cite{mohamad2015} a probabilistic decomposition was utilized to obtain analytical approximations for the full probability density function (pdf), including the tail structure, of systems subjected to parametric (or multiplicative) noise with correlated characteristics. In~\cite{tong15} the intermittent behavior of turbulent diffusion models with a mean gradient was rigorously quantified, while in~\cite{qi15} the capacity of imperfect models to capture intermittent behavior of turbulent systems was studied.  

The goal of this work is to  present a computational framework for   efficiently  quantifying  the statistical characteristics of extreme events. We focus on systems where uncertainty interacts with   system dynamics to produce~\emph{intermittent extreme events}, i.e. sporadically occurring large amplitude responses,  which give   rise to non-Gaussian statistics. Uncertainty could be due to    the initial data, parameters, or the dynamical system itself. The core idea of our probabilistic decomposition-synthesis method is the separation of   intermittent events from the background stochastic attractor,  in the spirit of the work done in~\cite{mohamad2015}, for low-dimensional systems. This decomposition  allows  us to apply~\emph{different} uncertainty quantification schemes for the two regimes (the background and the intermittent component). The background component, although potentially very high-dimensional, can be efficiently described by uncertainty quantification schemes that resolve low-order statistics. On the other hand, the intermittent component,  can be described in terms of a low-dimensional representation through a small number of  localized modes. The  probabilistic information from these two regimes  is synthesized according to a total probability decomposition argument, in order to approximate the   heavy-tailed probability distributions  for functionals of interest. Thus, the core of the approach relies on the assumption that heavy-tails are primarily due to the action of  intermittent instabilities, whereas the `main mass' of the probability density is due to the the background component. 

 To illustrate the method, we apply the developed framework on two systems that exhibit intermittently extreme responses and estimate their statistical distributions. The first example is a   nonlinear system of coupled random oscillators, which serves to illustrate the various steps of the method  in a simple  prototype,  where the mechanism behind intermittent instabilities  is easy to understand  although  the statistical characteristics  of the response are highly  non-trivial. The second example, is  a nonlinear envelope equation characterizing the propagation of unidirectional water waves in  deep water. The   benefits  of the method are  highlighted in this complex example, where we are able to capture the statistics at a fraction of the cost of direct numerical simulations. In both cases, we compare our estimates for the statistical distribution with direct Monte-Carlo results, and  illustrate   the  performance of the  approach.

The paper is structured as follows. In~\cref{sec:problem_setup} we describe the problem setup. In~\cref{sec:decomposition} we detail the various steps for  the proposed decomposition-synthesis method. \Cref{sec:rare_events} we    analyze  and detail the  various order reduction schemes for the statistical quantification for  rare events and the background attractor.  In~\cref{sec:osc_eg} we illustrate the method to the first example of a coupled, nonlinear system of random oscillators. In~\cref{sec:mnls}, we demonstrate the method on the second example to nonlinear  waves in  deep water.

\section{Problem setup}\label{sec:problem_setup}

Let $(\Omega,\mathcal{B},\prob)$ be a probability space, where $\Omega$ is the sample space with $\omega \in\Omega$ denoting an elementary event of the sample space, $\mathcal{B}$ the associated $\sigma$-algebra of the sample space, and $\prob$ a probability measure. (In the following, $\prob_X$ will denote the probability measure associated with a variable $X$ and $\pdf_X$ the associated  probability density function, pdf, if appropriate).

Let the \emph{dynamical system} of interest be governed by the following stochastic Partial Differential Equation (SPDE):
\begin{equation}
    \frac{\partial u(x,t)}{\partial t} = \mathcal N [ u(x,t); \omega], \quad x \in D,\;\; t \in [0,T], \; \; \omega \in \Omega,
    \label{eq:general_eq}
\end{equation}
where $\mathcal N$ is a general (nonlinear) differential operator with appropriate boundary conditions. We assume the initial state at $t=t_0$  is random and described by $u(x,t_0) = u_0(x;\omega)$. In what follows we  utilize a spatial inner product, denoted for two arbitrary functions $u(x)$ and $v(x)$ as $u \cdot v$.
Extreme events are meant in terms of a norm $\lVert \blank \rVert$ (e.g. the spatial supreme norm). In addition, the ensemble average of a random quantity $f(\omega)$ is   denoted by  $\bar f $.

Here we are interested in determining the statistical distribution for a \emph{quantity of interest} given by a functional of the solution $u(x,t)$ or as a solution of another dynamical system subjected to $u(x,t)$:
\begin{equation}\label{eq:quantity_interest}
    q=q[u(x,t)], \quad \text{or}  \quad   \frac{dq}{dt} = \mathcal M[q, u(x,t)].
\end{equation}Examples of such quantities could be properties of inertial tracers in turbulent flows, stresses for ocean structures subjected to water waves, or strains of mechanical components subjected to parametric and/or additive excitations. The computational cost associated with the estimation of the heavy-tailed statistics for such quantities is vast given the fact that the dynamical system~\eqref{eq:general_eq}  is characterized by the occurrence of rare events, which define the heavy-tail properties for $q$. Therefore, application of direct methods, such as Monte-Carlo simulations are prohibitively expensive for these problems.

\section{The probabilistic decomposition-synthesis method}\label{sec:decomposition}

We describe a systematic method to quantify the non-Gaussian response statistics (due to rare  events caused by intermittent instabilities) in a computationally efficient manner. More specifically, in this section we  first give an overview of the steps involved for the decomposition-synthesis method; more details are provided in the following section.

\subsection*{Decomposition and assumptions}

We assume that all rare event states due to internal instabilities, defined by the condition $\lVert u\rVert>\zeta,$ with $\zeta$ being the rare event threshold,   `live' in a low dimensional subspace $V_s$. We then decompose the response of the system as:
\begin{equation}\label{eq:decomp}
    u(x,t)=u_b(x,t)+u_r(x,t), \quad \text{with }\;   u_r=\Pi_{V_s}[u],\quad \text{if }\;\lVert u\rVert>\zeta,  \quad\text{and }\;  u_b=u-u_r,
\end{equation}
where $\Pi_{V_s}$ denotes the linear projection to the subspace $V_s$.
A similar decomposition onto a subspace of interest and a background component, but without taking into account the conditioning on rare events, has been utilized successfully for the uncertainty quantification and filtering of turbulent systems~\cite{sapsis_majda_mqgdo, Majda2014}. This conditional decomposition  will allow for the study of the two  components separately (but taking into account mutual interactions), using
\emph{different} uncertainty quantification methods that (i)  take into consideration
the possibly high-dimensional (broad spectrum) character of the stochastic background,
and (ii) the nonlinear and unstable character of   rare events.

\begin{figure}[tb]
    \centering
    \includegraphics{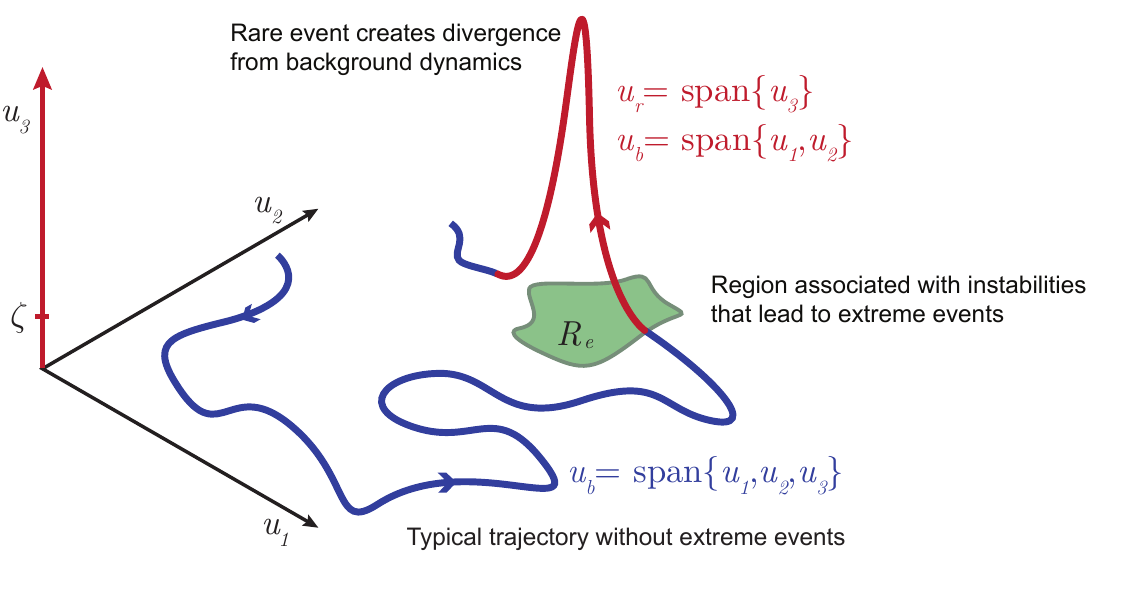}
    \caption{The conditional decomposition~\eqref{eq:decomp} partitions the system response only when a rare event occurs due to an instability. This happens when the state of the system enters the instability region $R_e$. In this example the subspace  associated with rare events due to instabilities is $V_s=\spn\{u_3\}$.}
    \label{fig:phaseDecomp}
\end{figure}

\textit{In this work we are interested in rare events that occur  due to transient instabilities}. To this end, we denote as $R_e$ the set of all background states that trigger an instability that   lead to a rare event. In~\cref{fig:phaseDecomp} we demonstrate the adopted decomposition in a three-dimensional system where the rare event subspace is defined by the linear span of $u_3$.  

The application of this decomposition onto a stochastic background and rare events relies on the  following assumptions:
\begin{enumerate}
\item The existence of intermittent events have negligible effects on the statistical characteristics of the stochastic attractor and can be ignored when analyzing  the  background state $u_b$.
\item Rare events are statistically independent from each other.
\item Rare events are characterized by low-dimensional dynamics.
\end{enumerate}

The first assumption  allows for the application of closure models or representation
methods that can deal with the high dimensional character of the stochastic background attractor. It expresses the property
that the rare events, although of large magnitude, are localized in time and space and can  induce only negligible modifications to the statistics of the background state.
The second assumption   follows from the rare character of   extreme events.
As for the third assumption,  related to the  low-dimensionality of the dynamics
of rare events,  follows naturally from their spatially or temporally localized
character. We emphasize  that all these assumptions do not imply any restrictions on the dimensionality of the stochastic background state.

\subsection*{Analysis of the various regimes}
The analysis of the two regimes will consist of the following steps:
    \begin{enumerate}
        \item \textbf{Order-reduction} in the subspace $V_s$ in order to model the rare event dynamics, expressed through $u_r.$  Then using the approximation   $u(x,t)\simeq u_r(x,t)$ we will compute the conditional pdf $\pdf(q\mid \lVert u \rVert>\zeta,u_{b}\in R_{e})$, under the condition that  an extreme event occurs due to an internal instability in $R_e$.
        \item \textbf{Quantification of the instability region} $R_e$ using the reduced-order model, by analyzing the conditions that lead to a rare event.
        \item \textbf{Description of the  background dynamics}, expressed through the statistics of $u_b$, which is not influenced by any internal instabilities in $R_e$.  Thus, when the response is dominated only by the background dynamics, we   have  $u(x,t)= u_b(x,t)$ and the pdf for the quantity of interest is given by  $\pdf(q\mid u=u_{b}).$  
        \item \textbf{Probability for rare events due to internal instabilities} $\prob(\lVert u \rVert>\zeta,u_{b}\in R_{e})$, which quantifies the total time/space that the response spends in the rare event regime due to the occurrence of instabilities.   
\end{enumerate}

\subsection*{Probabilistic Synthesis}

The next step of our technique is to probabilistically synthesize the information obtained from the previous analysis.  Using a total probability argument, in the spirit of~\cite{mohamad2015}, we  obtain the statistics for the quantity of interest  $q$   by
\begin{equation}\label{eq:probLaw}
\pdf(q) =  \pdf(q \mid {\left\Vert u \right\Vert>\zeta,u_{b}\in R_{e}} ) \prob_{r} +  \pdf(q \mid u=u_{b})\left( 1-\prob_{r}\right),
\end{equation}
where $\prob_{r}= \prob(\lVert u \rVert>\zeta, u_{b}\in R_{e}) $ is the probability of a rare event due to an instability.  The first term expresses the contribution of   rare events due to internal instabilities and is the heavy-tailed part of the distribution for $q$. The second term expresses the contribution of the background state and is the main    main probability mass in the pdf for $q$.

\begin{figure}[tb]
    \centering
    \includegraphics{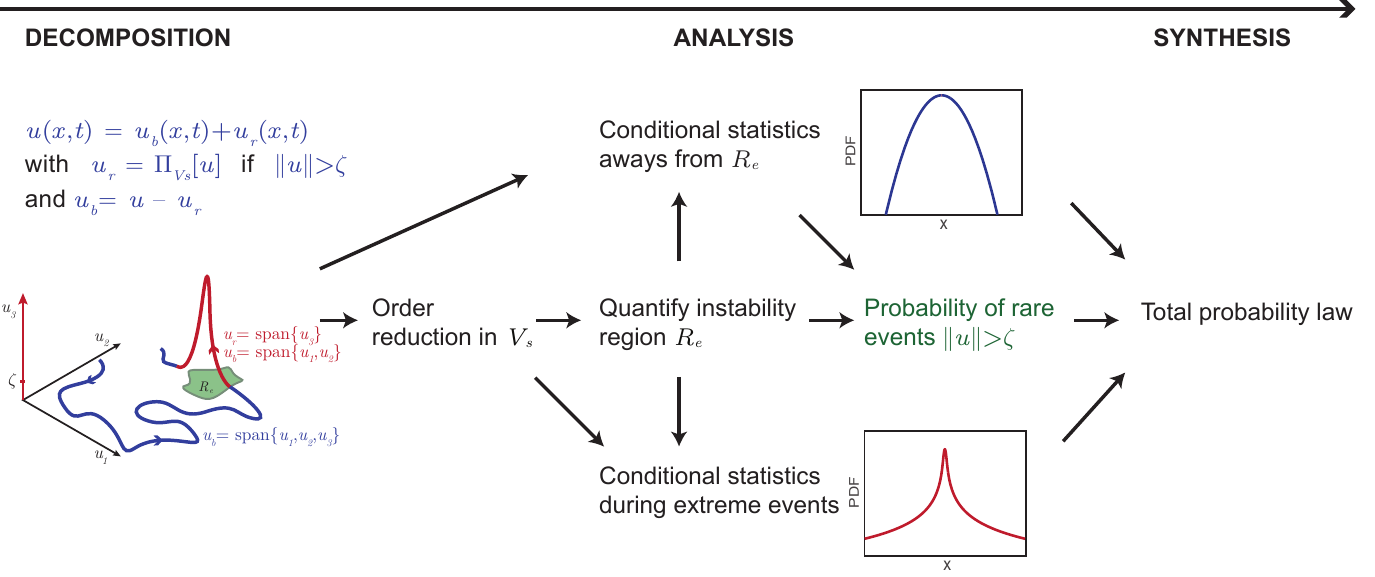}
    \caption{Outline of the steps  of the decomposition-synthesis method.}
    \label{fig:methodFlowChart}
\end{figure}

This total probability decomposition separates the full response  into the
conditionally extreme response and the conditionally background response,
weighted by their appropriate probabilities. The decomposition~\eqref{eq:probLaw} 
separates  statistical quantities according to the total probability law through conditioning on  \emph{dynamical regimes}. In this manner,   our approach connects the statistical quantities that we are interested in with important dynamical regimes that determine the dominant statistical features (e.g. a Gaussian core due to the background state and exponential like heavy-tails due to  intermittent bursts). An outline of all the steps involved is presented in~\cref{fig:methodFlowChart}.

For low-dimensional systems, this decomposition
can   lead to   analytical results for the probability distribution of
the response, which would otherwise be impossible to obtain from the dynamical
system and properties of the noise. For more complex systems, the primary
benefits are computational. We can resolve tail statistics with far fewer simulations
than direct Monte-Carlo simulation would require, since \textit{the decomposition allows for the evaluation of the tails by targeted simulations of rare events}, as opposed to the
random sampling of Monte-Carlo simulations. 

We emphasize that the expensive computation in the proposed algorithm is the conditional statistics for rare events $\pdf(q \mid {\lVert u \rVert>\zeta,u_{b}\in R_{e}} ).$  Once this has been obtained, it is simple to compute  rare event statistics for different configurations of the background state, since the other (non-rare) quantities are easy to compute multiple times. This contrasts with the Monte-Carlo method, where
sampling must be performed anew when noise and/or system parameters are changed.
Furthermore, with the proposed method the statistics of arbitrary functionals
of the response can be easily obtained with minimal additional computational expense, whereas Monte-Carlo simulations may be prohibitively expensive in this case.

\section{Detailed description of the analysis steps }\label{sec:rare_events}

We now provide the details for the analysis of each component starting from the reduced-order description of rare events and continuing with the statistical quantification of the background dynamics and the probability of rare events.

\subsection{Order reduction of the rare event dynamics}

The first  step for the effective description of the dynamics of rare events is the selection of a reduced-order basis or a reduced-order subspace denoted as $V_s$. These  events are spatially localized structures,   to this end we  employ  localized basis around a neighborhood of an arbitrary point $x_c$ at time $t_c$:
\begin{equation}
\hat {  v }_{ n } ( x-{ x }_{ c },t-t_c ) ,\quad n=1,\ldots,s,
\end{equation}
where $s$ is the dimension of the subspace. Note that the subspace  explicitly depends on the spatiotemporal location $(x_c, t_c)$ of the rare event, which is arbitrary. There are numerous methods with variable complexity that   can be utilized to choose or compute the subspace of rare events. The simplest choice in this case is a steady localized basis, such as Gabor basis or wavelets~\cite{cousins_sapsis}. However, if there is important spatial translation of the rare event during its lifetime then it may be beneficial to utilize adaptive methods for the evolution of the basis elements    $\hat{v}_{ n }$, such as the dynamically orthogonal field equations~\cite{SapsisLermusiaux09, sapsis11a}.

Apart from the selection of an appropriate set of basis functions, there is a variety of options for performing the order-reduction of the dynamics. Here we   discuss two different approaches for obtaining the local dynamics, which we   later illustrate through two specific problems.

\subsubsection{Projection of the dynamics around a zero background state}\label{proj1bk}
The simplest strategy to study the reduced-order dynamics of rare events is to perform a Galerkin projection ignoring the background state $u_b(x,t)$ for the evolution of the rare events inside the reduced-order subspace $V_s$. \textit{Such an assumption is valid if the background state plays a role only for the initial triggering of a nonlinear  instability}, while its small magnitude (compared with the intense local  responses associated with the   nonlinear dynamics in $V_{s}$) has very small influence on the reduced-order dynamics of the rare event.

Therefore, with the proposed order-reduction strategy we study rare events assuming that their evolution is isolated from the stochastic background. The coupling with the background state is introduced only through the background conditions at $t=t_c$, which is the initial condition for the reduced-order dynamics within $V_s$.
Based on this setup, we express the intermittent event in terms of the localized basis that span  the subspace $V_s$:
\begin{equation}\label{eq:projec2}
u_{ r }(x,t)=\sum _{ n=1 }^{ s } { a }_{ n } ( t )  \hat{v}_n(x -  x_c, t - t_c).
\end{equation}
Projecting the local (in the sense of $x_c$) dynamics of the full nonlinear equation on the subspace $V_s$, we obtain the following local dynamical system:
\begin{equation}\label{ic_proj1}
        \begin{aligned}
             \dot a_i &=\mathcal N \biggl[\,  \sum _{n=1}^{s}{a_n \hat v_n } ;\omega \biggr] \cdot {\hat v_i},\quad &i=1,\ldots,s, \\
            a_{i0} &=   (u_b\cdot\hat{v}_i )|_{ t=t_c },   &i=1,\ldots,s,
        \end{aligned}
\end{equation}where we again emphasize that the reduced-order system contains information about the background state only through the initial condition at $t=t_{c}$.
Through this reduced-order description we  can obtain the conditions (in terms of $a_{i0}$ and $\omega$) that define $R_{e}$, the domain of attraction to rare events.

\subsubsection{Projection of the dynamics around the stochastic background attractor}\label{red2}
A more accurate approach for quantifying localized instabilities is the formulation of reduced-order models around the background attractor, instead of a zero background state. This can be critical if the background state   not only triggers intermittent instabilities but also determines their evolution forward in time.

Here we  utilize the assumption that the presence of the localized instability $u_{r}$ does not have   important influence on the evolution of the background state $u_{b}$. We then express the full solution as in decomposition (\ref{eq:decomp}):
\begin{equation}
u(x,t)=\Pi _{V_s^{\bot}} [u(x,t) ]+\sum _{ n=1 }^{ s }{  a
_n } ( t )  \hat v_n (  x-x_c ,t-t_c  ),\\
\end{equation}
where the projection operator  onto the orthogonal complement is defined as:
\begin{equation}
\Pi_{V_s^\bot}[ u ]\equiv u-\sum_{ n=1 }^{ s }{ (u \cdot\ \hat{v} _{ n })  \hat{ v }_{ n }}.
\end{equation}
Using this setup we project the original equation on the basis elements $\hat{ v } _{ i }$ to obtain:
\begin{equation}
{ \dot  a   }_{ i } =\mathcal N\biggl[\Pi _{V_s^{\bot}} [u]+ \sum_{n=1}^s{  a_n  \hat  v_n } ;\omega \biggr] \cdot \hat
v_i - \Pi_{V_s^{\bot}}\biggl[\frac{\partial u}{\partial t}\biggr] \cdot\hat
{ v } _{ i }, \quad
i=1,\ldots,s,
\end{equation}
where the last term on the right hand side vanishes identically. Therefore, the localized dynamical system takes the form:
\begin{equation}
\label{ROE2}
\begin{aligned}
  \dot{a}_i &= \mathcal N\biggl[\Pi _{V^{\bot}_{s}} [u] +
\sum_{n=1 }^{ s }{ {a_n} \hat v_n }; \omega \biggr] \cdot \hat {v_i },\quad  &i=1,\ldots,s,\\
a_{i0}&=   (u_b \cdot \hat v_i  )|_{t=t_c},    &i=1,\ldots,s.
\end{aligned}
\end{equation}
Clearly, if we set the background state to zero, the   dynamical system above reduces to the formulation derived previously. The description of the background state is obtained by projecting the full equation in $V_s^\perp$ and taking into account that the evolution of $u_r$ has negligible effect on the projected dynamics of the background component. However, the infinite dimensional character of the dynamics in the orthogonal complement, $V_s^\perp$, makes it impractical to utilize the full equations for $\Pi _{V^{\bot}_{s}} [u]$. To this end, the reduced-order dynamical system (\ref{ROE2})  should be seen
as a starting point, where appropriate finite-dimensional truncations of
$\Pi _{V^{\bot}_{s}} [u]$ can be utilized that   capture the essential
effects of the stochastic background on  the evolution of the  localized
rare
events. Such an approach will be demonstrated in the problem involving water waves.

\subsection{Quantification of the instability region}\label{sec:domain_attraction}

To simplify the analysis we
  consider the case where the reduced-order dynamics have no important dependence
on $\omega$. We define $R_e$ as the set of background states for which the
reduced-order dynamics in $V_s$ posses at least one finite-time Lyapunov exponent that is positive.
More specifically, let the flow map $\phi^{t}_{t_{c}}:V_s\rightarrow V_s$
that maps any point $\mathbf a_0=\Pi _{V_{s}} [u]$ to its position at time time $t$ under
the effect of the dynamical flow (\ref{ic_proj1}):
\begin{equation}
\phi^{t}_{t_c}:V_s\rightarrow V_s,\quad \mathbf a_0 \mapsto \mathbf a(t,t_c,\mathbf
a_0;u_{b}).
\end{equation}
Note that this dynamical flow may also depend on the background state  $u_b \in V_s^\perp$. For each initial condition $\mathbf a_0$ we define the maximum finite-time
Lyapounov exponent over some finite-interval~$\tau$:
\begin{equation}
\lambda_{\tau}=\max _{i=1,\ldots,s}\log l_{i}\Bigl( \bigl[ \nabla_{\mathbf{a}_0}\phi^{t_c+\tau}_{t_{c}}
\bigr]^{*}\nabla_{\mathbf a_0}\phi^{t_c+\tau}_{t_{c}}\Bigr),
\end{equation}
where $  l_{i}$ denotes the eigenvalue of the Cauchy-Green tensor, which
is by definition symmetric and positive-definite. The domain of attraction
to rare events $R_e$  is then   defined as the set of background states
for which we have expansion in the reduced-order subspace:
\begin{equation}
R_{e}= \bigl\{ \mathbf a_0 \in V_{s},u_b \in V_s^\perp    \bigm\vert  \lambda_{\tau} ( \mathbf{a}_0
;u_{b}) >0 \bigr\}.
\end{equation}

\subsection{Description of the  background dynamics}
Here we discuss methods for the representation of the statistics of the background stochastic attractor. This refers to the part of the decomposition (\ref{eq:decomp})  $u_b$  for which instabilities have no role. There are numerous ways to approach the problem and here we review some of the techniques that can be used.

\subsubsection{Gaussian closure }
Based on our setup,  the stochastic background $u_b$ does not contain  intermittent events due to instabilities; as a consequence, it is reasonable to assume that its statistics can be approximated by a Gaussian distribution. This assumption can be the starting point for the application of closure schemes.

For systems  that are characterized by a stable mean state $\overbar  u $, the finite variance of the steady state attractor is caused by the   external stochastic excitation (see~\cite{sapsis_majda_qgdo} for more details). In this case partial linearization of the dynamics or a Gaussian closure of the infinite system of moment equations (see e.g.~\cite{epstein69})    is  an effective option to capture the conditional statistics of the system in the state $R_e^{c}$, where we have no rare events occurring. For a such situation the governing equations can be linearized to give
\begin{equation}
\frac{\partial u_b}{\partial t} = \mathcal N [ \overbar u;\omega]+ \mathcal L_{\overbar{u}}  u_b, \quad
 \omega \in \Omega,
\end{equation}
where the first term on the right-hand side contains deterministic and stochastic noise terms that are independent of the state $u_b$, while the second term denotes the linearization of the system,
\begin{equation}
 \mathcal L_ {\overbar u} = \frac{\delta \mathcal N}{\delta u}\bigg|_{\overbar u}
\end{equation}
around the mean $\overbar u,$  where all the eigenvalues have a negative real part (stable mean state). For such system it is straightforward to formulate the  second order moment equations and obtain an expression for the Gaussian measure that characterizes the statistics of the background attractor.

Such an approach will not be effective if the system under consideration has persistent instabilities that lead to nonlinear energy transfers between modes, i.e. the mean state is not stable. For such cases other methods may be used to obtain  representations of the background attractor statistics, such mean stochastic models~\cite{sapsis_majda_tur} or quasi-linear Gaussian closures~\cite{sapsis_majda_mqg,sapsis_majda_tur}.

\subsubsection{Gaussian representations based on conditional Monte-Carlo simulation}\label{sec:gausconditional}

A more direct approach involves the numerical simulation of the system and the conditional sampling of the  second-order statistics of the background state using the representation (\ref{eq:decomp}).
It is also often the case that the spectrum of the full random field has been measured or estimated. Such an approach is typical, for example, in water waves or other geophysical systems. 

\subsubsection{Analytical approach using the Fokker-Planck-Kolmogorov equation}\label{sec:decomp_analyticalfpk}

For systems having background states that can be modeled by  dynamical systems of a  special form, analytical  descriptions of their stationary probability measure may be available. In particular, for systems excited by white noise one can formulate the corresponding Fokker-Planck-Kolmogorov (FPK) equation defining the evolution of the state pdf~\cite{Sobczyk91,Soong_Grigoriou93}. For the special case of vibrational systems possessing a Hamiltonian structure perturbed by  additive and/or parametric white noise under linear damping, for special conditions, the stationary measure has an explicit form   in terms of the Hamiltonian of the system~\cite{soize88,soize94}. Furthermore, in~\cite{caughey82,zhu90,wang00} an analytical approach is utilized for determining the stationary pdf of more generic vibrational systems, where the steady FPK equation is solved by splitting to simpler partial differential equations.

To demonstrate the analytical approach, we consider the case of a background state described by a collection of decoupled vibrational modes, each governed by an equation of the following form:
\begin{equation}
\ddot u + \delta \dot u + \nabla V(u) =\sigma\dot W(t;\omega).
\end{equation}
Then, the pdf in the statistical steady state  will be given in terms of the Hamiltonian \(\mathbb{H}(u,\dot u)\) of the system as~\cite{Sobczyk91}:
\begin{equation}
\pdf_{u\dot u}(u,\dot u)=C\exp\left(-\frac{2\delta}{\sigma^2}\mathbb{H}(u,\dot u)\right),
\end{equation}
 where $\mathbb{H}(u,\dot u)=\frac{1}{2}\dot u^2 + V(u)$, and $C$ is a normalization
constant. We   utilize this approach in the first example involving a nonlinear vibrational system.

\subsection{Probability for rare events $\prob(\lVert u \rVert>\zeta,u_{b}\in R_e)$}

Having determined the conditional statistics of the background dynamics, the final step is to determine the probability of   rare transitions, i.e. the likelihood of  $\lVert u \rVert>\zeta$ when an instability occurs, i.e. $u_{b}\in
R_e$.  This will be defined for an arbitrary point in space, $x_0$, through the integral:
\begin{equation}
\prob(\lVert u \rVert>\zeta,u_{b}\in
R_e)=\frac{1}{T}\int\limits_{t \in T} \mathbbm{1}(\lVert u(x_{0},t) \rVert>\zeta,u_{b}\in R_e)\, dt,
\end{equation} where  $ \mathbbm{1}$ is the indicator function. This integral measures the duration of rare events compared with the overall time interval.

Note that the above probability is not directly equal to the probability of the background state crossing into  the instability region $R_e$, which is a condition just for the occurrence of a rare event and does not contain information regarding the duration of the rare event, which can last even if the system background has moved outside the instability region. 

The probability that we are interested in will be found using information obtained from the reduced-order model
developed previously. In particular, using the reduced-order model we will obtain the temporal extent $T_{e}(u)$  for each rare event that corresponds to any unstable point $u \in R_e$.  Note that for states $u$ not associated with instabilities $T_{e}(u)=0$. 

Then the
probability of rare transitions due to instabilities will be approximated by
\begin{equation}\label{eq:rare_trans}
\mathbb{P}(\left\Vert u \right\Vert>\zeta,u_{b}\in
R_e)=\frac{1}{T}\int\limits_{u } \! T_{e}\left(u\right)\pdf_{u_b}(u)\,du ,
\end{equation}
where $\pdf_{u_b}=\pdf(u \ \mid u=u_{b})$ is the conditional pdf of the background state, which contains  no information on rare events.


\section{Nonlinear  system of coupled  oscillators}\label{sec:osc_eg}
\begin{figure}[htb]
        \centering
        \includegraphics{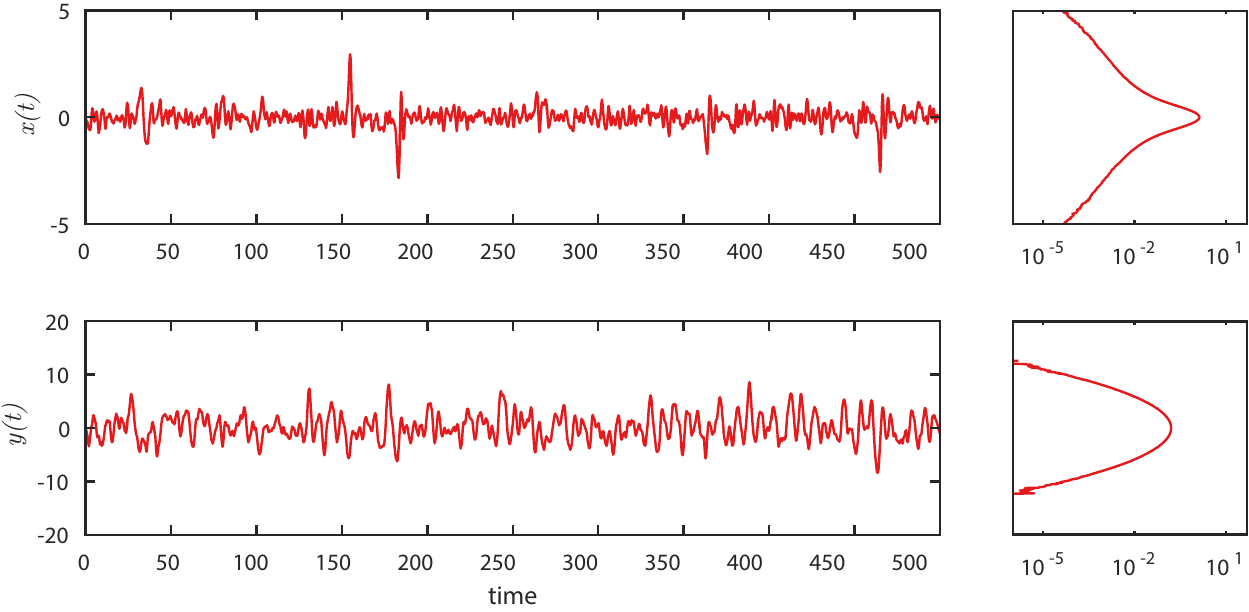}
        \caption{Realization of the intermittent variable $x(t)$ (top, left) with large amplitude   bursts and a fat-tailed equilibrium pdf (top, right), alongside the exciting signal  $y(t)$ (bottom, left) with a Gaussian response pdf (bottom, right).}
        \label{fig:oscil_real}
\end{figure}

We first illustrate the decomposition-synthesis method on   a  two-degree-of-freedom system composed of a linear oscillator  coupled quadratically to a nonlinear oscillator with cubic stiffness,   both under the action of white-noise excitation. The system is given by
\begin{equation}\label{eq:osc_system}
        \begin{aligned}
        &\dd{x}{t} +  c_x \d{x}{t} + \bigl(k_x + a_x y(t) \bigr)x  + s x^3  = \sigma_{f_{x}} \dot W_x , \\
        &\dd{y}{t} + c_y \d{y}{t} + k_y y  = \sigma_{f_{y}} \dot W_y , 
        \end{aligned}
\end{equation}
where $x,y$ are the two state variables, $c_x,c_y > 0$ are the damping parameters, $k_x,k_y>0$ are   the linear stiffness parameters, and $s\neq0$ is the nonlinearity parameter. The constants $\sigma_{f_x},\sigma_{f_y} > 0$ are the noise intensities, and $W_x,W_y$ are two independent scalar Wiener processes. With $a_x\neq 0$, the system (\ref{eq:osc_system}) possesses regimes with  intermittent instabilities  that lead to fat-tailed equilibrium pdfs in the variable $x$, whereas the variable $y$ converges to a  Gaussian   pdf (see~\cref{fig:oscil_real}). This system is introduced such that  $y$  can trigger an intermittent response in $x$  through a large deviation from its mean value. 

System~\eqref{eq:osc_system}      is prototypical of the action of intermittency in      more complex systems   where similar interactions are at play between system modes, e.g. internal instabilities   associated with transfers of energy between modes in turbulence, buckling of beams and plates under stochastic excitation, ship rolling under parametric stochastic resonance, just to mention a few. Although  system~\eqref{eq:osc_system} is   low-dimensional, quantifying the stationary pdf for the variable $x$ is  challenging. For one, the  finite time-correlated parametric excitation term  due to  $y$   on the variable $x$ (which is the mechanism behind  transient instabilities), makes application of the Fokker-Planck equation, unpractical and computationally prohibitive (the resulting equations become high-dimensional). Moreover,  there is a  nonlinear restoring force ($s\neq0$) that has to be taken into consideration; this nonlinear  term   has a significant impact on the pdf of $x$.   

Besides its prototypical character, part of the motivation behind the current example is to apply our proposed method to a system where the mechanism behind rare events is transparent. As a result, we do  not need to apply the  order-reduction schemes   described in~\cref{sec:rare_events} for the rare event component (this will be done in the next application).  This will allow for the demonstration of the   decomposition-synthesis procedure   in a   manner that can be adapted to different problems that share similar characteristics.

\subsection{Decomposition and instability region $R_e$}
In this case the decomposition step is trivial since all the rare events involve the variable $x$, thus $V_s=\spn\{ (1,0)^{T} \}$. In particular, intermittency in the variable $x$ is a consequence of $\kappa(t) \coloneqq k_x + a_x y(t)$ switching signs from positive to negative values, during which $x(t)$ transitions from its regular  response to a domain where the likelihood of an instability is high or guaranteed.  This switching in $\kappa(t)$ is the triggering mechanism behind localized instabilities in the variable $x(t)$. Therefore, we define the instability region\begin{equation}
R_{e}= \{ y  \mid  \kappa =k_x + a_x y<0 \}.
\end{equation}

\subsection{Conditional statistics of rare events}\label{subsec:osc_domattraction}

The intensity of the rare events depends both of the average magnitude of $\kappa$ when this becomes negative, but also on the duration of the   downcrossing. Therefore, we will choose the rare event description presented in section \ref{red2}, where the background state is taken into account for the evolution of the rare events. We choose to parameterize the instability region $R_e$ with the   average amplitude $\alpha$ and average duration $\xi$ of each downcrossing event of $\kappa$, which expresses the background state information.   

\begin{figure}[tb]
        \centering
        \includegraphics{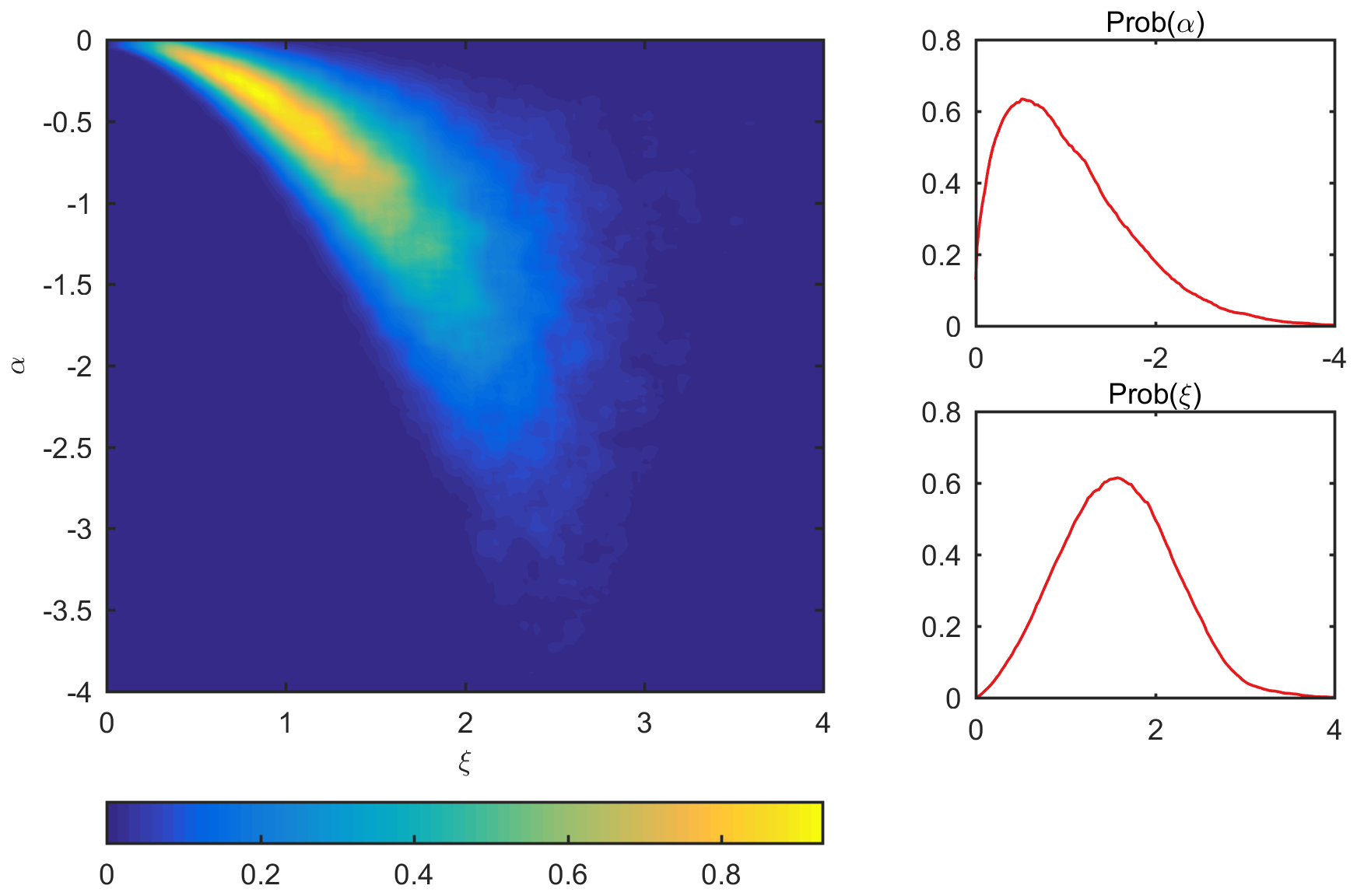}
        \caption{Joint density $\pdf_{\alpha \xi}(\alpha,\xi)$ (left), alongside the marginalized densities for   amplitude $\alpha$ (top right) and length scale $\xi$ (bottom right) for $c_y=0.5,$ $k_y=1$, $\sigma_{f_y}=2.5$.}
        \label{fig:oscil_PAL}
\end{figure}

We compute a few realization of the background state \begin{equation}
        \dd{y}{t} + c_y \d{y}{t} + k_y y  = \sigma_{f_y} \dot W_y,
\end{equation}
and for each realization  we identify all the regions where $\kappa(t)  = k_x + a_x y(t) < 0$. For each of  these events, that is, a zero downcrossing followed by a subsequent zero upcrossing, we take the duration of time that $\kappa<0$  as  $\xi$ and assign a characteristic amplitude $\alpha$ by taking the  mean of $\kappa$ over the duration that $\kappa<0$ . Performing this procedure for all realizations gives us a set of samples of $(\alpha,\xi)$. Since we are interested in the probability at a given temporal location  we have an event $\kappa < 0$ with duration $\xi$ and amplitude $\alpha$, we scale the amplitudes in the resulting histogram  by their corresponding durations $\xi$ in order to correctly weight the samples. This gives $\pdf_{\alpha\xi}(\alpha,\xi \mid R_e)$: the probability of finding a rare event characterized by an excitation duration $\xi$ and amplitude $\alpha$. We display this pdf in~\cref{fig:oscil_PAL}.

Next, we proceed with the computation of the conditional statistics for rare events, $\pdf_x(x \mid \alpha,\xi)$.
 This will be based on a few simulation of the equation determining $x$ taking into account the background state (expressed through $(\alpha,\xi)$). More specifically, for a given $\alpha_i,\xi_i$ we evolve the rare event dynamics
 according to
\begin{equation}\label{eq:osc_extremerun}
        \dd{x}{t} +  c_x \d{x}{t} + \tilde \kappa(t;\alpha_i,\xi_i) x  + s x^3 
= \sigma_{f_x} \dot W_x,
\end{equation}
where $\tilde\kappa(t;\alpha_i,\xi_i)$ is a  local representation of the function  $\kappa(t) = k_x + a_x y(t)$  in  the critical regime with
the same mean amplitude $\alpha_i$ (over $0\leq t\leq \xi_i$)  and duration
$\xi_i$. For simplicity, we choose the following function to approximate $\kappa$:
\begin{equation}
        \tilde\kappa(t,\alpha_i,\xi_i) = 
        \begin{dcases}
                \frac{\pi\alpha_i}{2} \sin( \pi t /\xi_i), \quad &\text{for }
0 \le t \le \xi_i,\\
                k_x,\quad &\text{otherwise}.
        \end{dcases}
\end{equation}
Initial conditions   for $x$ are chosen from the background state,  which we will describe later. In this regime,   we are only interested
in extreme responses for which $\left\vert x(t) \right\vert>\zeta=\sigma_{b}$,
where $\sigma_{b}^2=\overbar{({x}|_{\kappa>0})^2}$ is the variance of the background stochastic
state. To perform the computation we simulate an instability
according to (\ref{eq:osc_extremerun})   for
a length of time that is long enough to ensure that   $x(t)$ returns back to the
 background regime, where $\left\vert x(t) \right\vert<\zeta$,  and  sample only the points before the response relaxes
back, i.e. the points for which $\left\vert x(t) \right\vert>\zeta$.
Following
this procedure we obtain the quantity $\pdf_x(x \mid \alpha,\xi)$.

\subsection{Conditional statistics of background dynamics}
\label{sec:osc_backgroundattractor}
Next we determine the statistical characteristics of the background attractor
$\pdf_x(x \mid R_e^{c})$. In this regime, by definition,  we have no rare
events, and we  choose to derive the statistical distribution    analytically.
This quantity could also be obtained more directly  through  the conditional
Monte-Carlo approach described in~\cref{sec:gausconditional}  by constraining
the sampling to the regime $\kappa=k_x + a_x y(t) > 0$.

Since  $\kappa > 0$ in  $R_e^{c}$, we choose to approximate $\kappa(t)$  by its conditionally average value  $\overbar{\kappa}|_{\kappa>0}$,
\begin{equation}\label{eq:osc_x_aved}
         \dd{x}{t} +  c_x \d{x}{t} + \overbar{\kappa}|_{\kappa>0} x  + s
x^3  = \sigma_{f_{x}} \dot W_x.
\end{equation}
The steady state pdf can be found
by   solution of the associated  FPK equation:
\begin{equation}
        \pdf_{x\dot x}(q,\dot q \mid R_e^{c}) = C \exp\biggl( - \frac{2c_x}{\sigma_{f_x}^2}
\mathbb H(q,\dot q) \biggr), \quad\text{where } \;\;  \mathbb H(q,\dot q) = \frac{1}{2}\dot
q^2 + \frac{1}{2} \overbar{\kappa}|_{\kappa>0} q^2 + \frac{1}{4}s q^4,
\end{equation}
and $C$ is a normalization constant. Marginalizing out $\dot x$ in the equation
above, gives the  conditional pdf for $x(t)$ in $R_e^{c}$:
\begin{equation}\label{eq:osc_stablepdf}
        \pdf_x(q \mid R_e^{c}) = C \exp\biggl[ - \frac{2c_x}{\sigma_{f_x}^2}
\biggl(\frac{1}{2} \overbar{\kappa}|_{\kappa>0} q^2 + \frac{1}{4}s q^4\biggr)
\biggr],
\end{equation}
where $C$ is the normalization constant.

To determine $\overbar{\kappa}|_{\kappa>0}$ we utilize the steady state pdf
of $y(t)$  which is    Gaussian distributed:
\begin{equation}\label{eq:yresponsepdf}
        \pdf_y(q) =  \sqrt{\frac{c_y k_y }{\pi \sigma_{f_y}^2}} \exp\biggl( -
\frac{c_y k_y}{\sigma_{f_y}^2} q^2\biggr).
\end{equation}
with variance $\sigma_y^2 = \sigma_{f_y}^2/(2 c_y  k_y)$. Therefore, we obtain
\begin{equation}
        \overbar{\kappa}|_{\kappa>0} = k_x +   a_x\sigma \frac{\phi\bigl(-\frac{k_x}{
a_x\sigma_y} \bigr)}{1-\Phi\bigl(-\frac{k_x}{ a_x\sigma_y}\bigr)},
\end{equation}
where $\phi(\blank)$ is the normal probability distribution function and
$\Phi(\blank)$ the normal cumulative distribution function.

\subsection{Probability for rare events}\label{sec:osc_liklihood}

We now proceed to compute the probability of rare events due to instabilities in $R_e$. This can be found by analyzing the duration of the rare transitions using~\eqref{eq:osc_extremerun}. Denoting by  $T_e(\alpha_i,\xi_i)$ the duration of the rare event corresponding to an instability in $R_e$ of average magnitude $\alpha_i$ and duration $\xi_i$, according to~\eqref{eq:rare_trans},  we have that
\begin{equation}
\label{prob_dur_osc}
\mathbb{P}(x>\zeta,y\in R_e)=\frac{1}{T} \int T_{e}\left(\alpha, \xi \right) \pdf_{\alpha \xi}(\alpha, \xi \mid \kappa < 0) \mathbb{P} (\kappa<0)\, d\alpha d\xi,
\end{equation}
where  
\begin{equation}
        \prob(\kappa<0) = \prob(y < -k_x/a_x) = \Phi \biggl( -\frac{k_x}{ a_x\sigma}\biggr).
\end{equation}
The probability~\eqref{prob_dur_osc} can  be approximated using the analytical argument presented~\cite{mohamad2015}, where the typical temporal duration of
the growth and relaxation phase of a rare event are examined. More specifically,   consider a single  representative extreme response with an average growth $\overbar{\Lambda}_+$ and decay rate $\overbar{\Lambda}_-$. During the growth phase we have that 
\begin{equation}\label{eq:osc_typicalgrowth}
        u_p = u_0 e^{\overbar{\Lambda}_+ T_{\kappa<0}},
\end{equation}
where $T_{\kappa<0}$ is the temporal duration of the growth event and $u_p$ is the peak value of the response and $u_0$ an arbitrary initial condition. Similarly, over the decay phase we have 
\begin{equation}\label{eq:osc_typicaldecay}
        u_0 = u_p e^{-\overbar{\Lambda} - T_\text{decay}},
\end{equation}
and by connecting these~\cref{eq:osc_typicalgrowth,eq:osc_typicaldecay}, 
\begin{equation}
        \frac{T_\text{decay}}{T_{\kappa<0}} = \frac{\overbar{\Lambda}_+}{\overbar{\Lambda}_-}.
\end{equation}
The  average duration of an extreme transition  is   then given by  $T_{e} = (1 + \overbar{\Lambda}_+/{\overbar{\Lambda}_-}) T_{\kappa<0}$. Therefore, by dividing over  the total time  duration $T$ we have 
\begin{equation}
        \mathbb{P}(x>\zeta,y\in
R_e)\simeq\  \biggl(1 + \frac{\overbar{\Lambda}_+}{\overbar{\Lambda}_-}\biggr) \prob(\kappa<0). 
\end{equation}To leading order (neglecting the nonlinear restoring term) $\Lambda_+  = \sqrt{- \kappa|_{\kappa<0}},$ since dissipation and additive forcing have a small role during the growth phase of an extreme response. Therefore, the average growth exponent is   $\overbar{\Lambda}_+ = \expec\bigl(\sqrt{-\kappa|_{\kappa<0}}\bigr),$ which is straightforward  to compute using $\pdf_{\alpha\xi}$. Similarly, we find that during the decay phase is $\Lambda_- = c/2$, and thus  $\overbar{\Lambda}_- = c/2$.

\subsection{Synthesis and comparison to Monte-Carlo simulations}

We have now determined  all the components required to construct the distribution of the response according to the decomposition-synthesis method. We synthesize  the results of the previous sections by the total probability law:
\begin{equation}\label{eq:osc_decompfinalresult}
\pdf_x(q) = \pdf_x(q \mid R_e^{c})\left( 1-\mathbb{P}_r \right) + \prob_{r}\iint \pdf_x(q \mid \alpha,\xi) \pdf_{\alpha \xi}(\alpha,\xi)\, d\alpha d\xi,
\end{equation}
where $\mathbb{P}_r=\mathbb{P}(x>\zeta,y\in R_e)$.

\begin{figure}[bt]
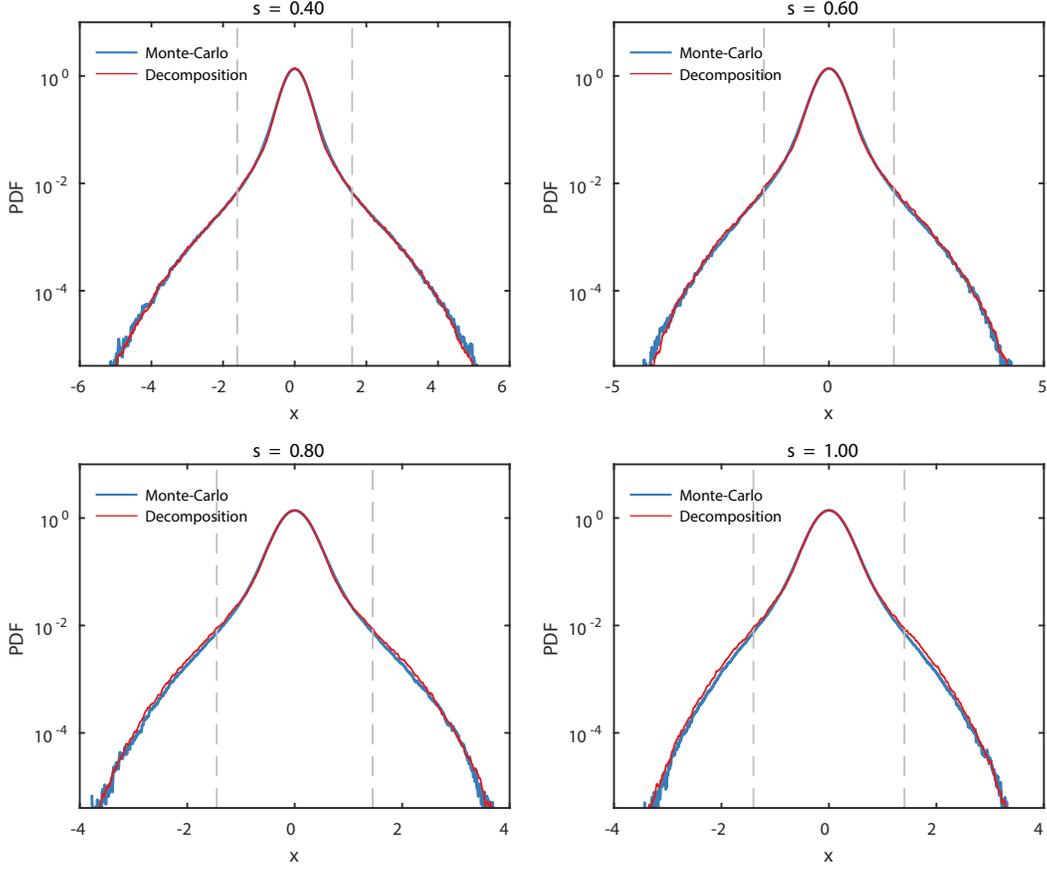

        \centering
        \includegraphics{{{response-pdf_xParams-c=1.00_m=4.00_s=0.40_sigx=0.75_yParams-b=0.50_k=1.00_sigy=2.50}}}
        \includegraphics{{{response-pdf_xParams-c=1.00_m=4.00_s=0.60_sigx=0.75_yParams-b=0.50_k=1.00_sigy=2.50}}}
        \includegraphics{{{response-pdf_xParams-c=1.00_m=4.00_s=0.80_sigx=0.75_yParams-b=0.50_k=1.00_sigy=2.50}}}
        \includegraphics{{{response-pdf_xParams-c=1.00_m=4.00_s=1.00_sigx=0.75_yParams-b=0.50_k=1.00_sigy=2.50}}}
        \caption{Probability distribution function approximation with the decomposition-synthesis method for the variable $x(t)$ for the nonlinear system of coupled oscillators~\eqref{eq:osc_system}. The blue line denotes the Monte Carlo simulation and the red line denotes
the approximation  by the decomposition-synthesis procedure.  The vertical dashed lines denotes $4$ standard deviations of $x(t)$. Parameters  are:  $c_x =1,$ $c_y=0.5,$ $k_x = 4,$ $k_y=1$, $\sigma_x = 0.75,$ $\sigma_y= 2.5$.} 
        \label{fig:osc_pdfresults}
\end{figure}

In~\cref{fig:osc_pdfresults} we compare the decomposition-synthesis results for the equilibrium pdf of $x(t)$ alongside the `true' density from Monte Carlo simulations,  for increasing values of the cubic stiffness parameter for a hardening spring $s>0$. Monte-Carlo results are computed using $5000$ realizations of (\ref{eq:osc_system}), integrated using the Euler–Maruyama method with time step $\Delta t = 0.002$ to $T = 500$ time units, discarding the first $t = 60$ data  to ensure a statistical steady state. Overall, we have very good quantitative agreement for both the tails and the core of the distribution  between the decomposition-synthesis approach  and the   Monte Carlo results.  For larger values of the nonlinearity parameter $s$, tail values are suppressed due to larger restoring forces. We see that this effect is also accurately captured for increasing $s$ values in the decomposition-synthesis approach. Here we emphasize the non-uniform decay character of
the tails which can still be captured very accurately. In particular we have favorable comparison for all three   qualitatively different  regimes: the core of the   distribution, the  exponential like heavy-tails at extreme values,  and the subsequent sub-exponential decay at very extreme values of distribution (where nonlinearity is very important).

We point out that computing the response pdf via the decomposition-synthesis for different background parameters $c_x,s,\sigma_x$ parameters is extremely cheap since   $\pdf_{\alpha \xi}(\alpha, \xi),$ which involves the rare event dynamics, remains fixed.  Also, if we are interested in the response pdf for different $k_x,a_x$,  this also has minimal computational cost because  we can store rare event realizations of $y$ and use them to determine $\pdf_{\alpha \xi}(\alpha, \xi)$, as required. Moreover, the computation to determine $\pdf_{\alpha \xi}(\alpha, \xi)$  is easy and fast, since we do not need a large number of realizations  to give good results when used in (\ref{eq:osc_decompfinalresult}) and  only requires the simulation of the background variable $y(t)$.

\section{Unidirectional nonlinear  water waves}\label{sec:mnls}

\begin{figure}[b]
        \centering
        \includegraphics{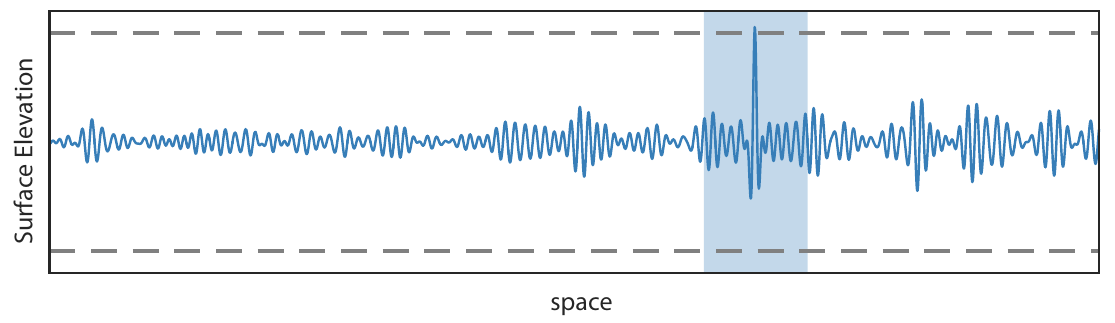}
        \caption{Example large amplitude freak wave formed due to nonlinear interactions.}
        \label{fig:exampleFreakWave}
\end{figure}

The second application   involves the problem of local extremes for nonlinear dispersive water waves. In particular, our goal  is to quantify the non-Gaussian, heavy-tailed distribution  of the local wave field maxima  in a nonlinear envelope equation characterizing the propagation of unidirectional water waves. This example has many ocean engineering applications, since quantifying  extreme water waves   is critical   for ocean structures and naval operations  due their catastrophic consequences. Indeed, extreme waves, termed  freak or rogue waves, have caused considerable damage to ships, oil rigs, and human life ~\cite{Dysthe08, Olagnon2005, Kharif2003, trulsen1996}. A wave is termed a freak if its crest-to-trough height exceeds twice the significant wave height $H_s$, with $H_s$ being equal to \emph{four times} the standard deviation of the surface elevation (\cref{fig:exampleFreakWave}). Therefore, such  waves are extreme responses that `live' in the tails of the distribution of the wave elevation. 

We consider waves on the surface of a fluid of infinite depth and work with approximate equations that govern the dynamics of the wave envelope. In particular, the evolution of a unidirectional, narrow-banded wave field  is described well by the modified Nonlinear Schrodinger (MNLS) equation of Dysthe~\cite{dysthe1979},  a high order approximation of the fully nonlinear model:
\begin{equation}        \label{eq:MNLS}
  \frac{\partial u}{\partial t} +\frac{1}{2}\frac{\partial u}{\partial x} + \frac{i}{8} \frac{\partial^2 u}{\partial x^2} - \frac{1}{16} \frac{\partial^3 u}{\partial x^3} + \frac{i}{2} |u|^2 u + \frac{3}{2} |u|^2 \frac{\partial u}{\partial x} + \frac{1}{4} u^2 \frac{\partial u^*}{\partial x} + i u \frac{\partial \phi}{\partial x} \bigg|_{z=0} = 0,
\end{equation}
where $x$ is space, $t$ is time, and $u$ is the envelope of the modulated carrier wave. The velocity potential $\phi$ at the surface may be expressed explicitly in terms of $u$, giving $\partial \phi/\partial x |_{z=0} = - \mathcal F^{-1}\bigl[ \abs{k} \mathcal F\bigl[\abs{u}^2\bigr] \bigr]/2$, where $\mathcal F$ is the Fourier transform. \Cref{eq:MNLS} has been nondimensionalized with $x = k_0 \tilde x$, $t = \omega_0 \tilde t$, $u = k_0 \tilde u$, where $\tilde x$, $\tilde t$, $\tilde u$  are physical space, time, and envelope variables, with $k_0$ the dominant spatial frequency of the surface and $\omega_0  = \sqrt{ g k_0 }$.  To leading order,  the   nondimensionalized surface elevation around the undistributed level is given by  $\eta(x,t) = \real[ u(x,t) e^{i\left(x-  t\right)} ]$. The local maxima of the surface elevation is described by the modulus of the envelope $|u|$, which is the quantity of interest in this example.

We consider a wave field with with initial characteristics given by
\begin{equation}
        u(x,0) = \sum_{k = -N/2+1}^{N/2} \sqrt{2 \Delta_k F(k\Delta_k)} e^{i(\omega_k x + \xi_k)}, \quad  F(k) = \frac{\epsilon^2}{\sigma \sqrt{2\pi}} e^{\frac{-k^2}{2\sigma^2}}
        \label{eq:gaussSpec}
\end{equation}
where $\xi_k$ are independent, uniformly distributed random phases between $0$ and $2\pi$. 

If the Benjamin-Feir Index (BFI), the ratio of steepness to bandwidth $\epsilon / \sigma$ is large enough, we have important probability for the occurrence of rare events due to nonlinear focusing. Such nonlinear focusing is triggered by the energy localization over a specific region, which is the result of 
phase differences between the various harmonics~\cite{cousinsSapsis2015_JFM} in the stochastic background $u_b$. These relative phases continuously change primarily due to the effect of linear dispersion and as the BFI increases there is a higher probability for them to result in important energy localization for $u_b$ and a subsequent nonlinear focusing event $u_r$. 

It has been shown~\cite{cousinsSapsis2015_PRE} that for unidirectional waves the occurrence of nonlinear focusing of an arbitrary  wavegroup (formed due to linear dispersion) in $u_b$ is controlled by the wavegroup amplitude $A$ and its lengthscale $L$. This fact leads us to the adoption of the wavegroup characteristics  $A$ and $L$ as a way to parameterize  the instability region, i.e. the region where nonlinear focusing occurs.

\subsection{Rare events subspace and  instability region $R_e$}\label{sec:mnls_rareventspace}

 We  describe the  dynamics of focusing wavegroups through a reduced
order subspace, $V_s$, obtained using a proper orthogonal decomposition~\cite{Holmes_et_al96} of
the focusing wavegroups under MNLS dynamics. The proper orthogonal decomposition is appropriate  in this case  as  the order reduction scheme for focusing wavegroups,  since we do not have important spatial translations of the focusing waves and this allows us to capture the dynamics with just a few modes. 

To capture the variations in the dynamics for different wavegroup lengthscales and amplitudes, we choose $n = 8$ sets of simulations of the MNLS for various $(A,L)$  that undergo nonlinear focusing. For each simulation we take snapshots $u_{s}(x,t_k; A_i,L_i)$ in time, where for each simulation we ensure that the snapshots are capturing the dominant focusing action. Stacking the snapshots for all the   simulations   gives the matrix $X$ 
\begin{equation}
        X = \bigl[ u_s(x,t_1; A_1,L_1) \;\; u_s(x,t_2;A_2,L_2)  \;\;\ldots  \;\; u_s(x,t_m; A_n,L_n)\bigr],
\end{equation}
where $m$ is total number of snapshots. We use the method of snapshots~\cite{Sirovich87} to determine the orthonormal POD modes, by solving the $m\times m$ eigenvalue problem $X^T X U = U \Lambda$, for the modes $V = X U \Lambda^{-1/2}$.  

\begin{figure}[tb]
        \centering
        \includegraphics[width=0.9\textwidth]{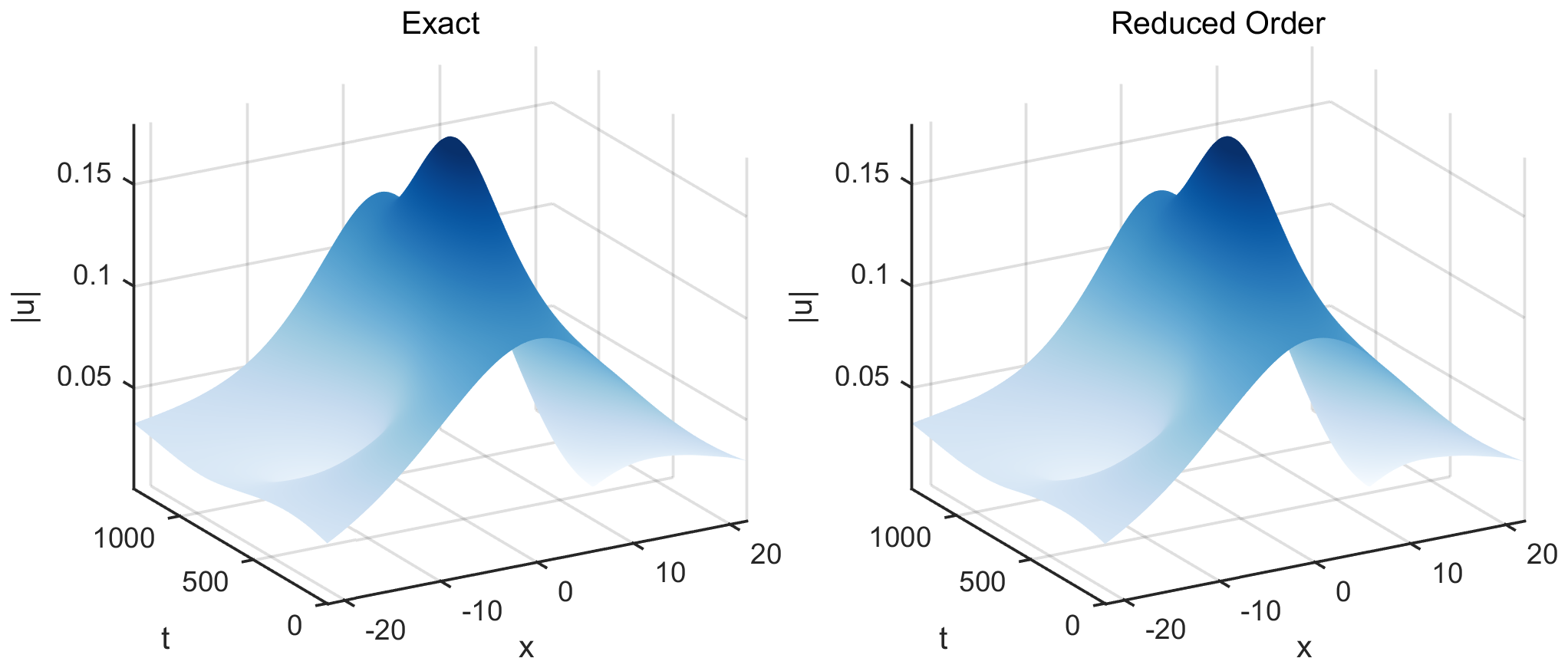} 
        \caption{Simulation of an extreme wave group for  $A = 0.11,$ $L = 11$, comparing the exact MNLS (left) and the reduced order model (right) with $14$ modes.}
\label{fig:comp}
\end{figure}

By this  procedure we obtain the local basis  $\hat v_{n}(x,t)$ and represent  $u_r$ as  
\begin{equation}\label{eq:mnls_pod}
u_{ r }(x,t)=\sum _{n=1}^{s} {a}_{n} (t)  \hat{v}_n(x -  x_c, t - t_c),
\end{equation}
which is conditional upon a background state at $(x_c,t_c)$. We used the projection of the full MNLS around a zero background state as described in~\cref{proj1bk}. We found that a projection upon $s = 14$ modes approximated the dynamics of $V_s$ well across a range of  $L$ and  $A$ (fig. \ref{fig:comp}).

\begin{figure}[htb]
\centering
\includegraphics[width=0.8\textwidth]{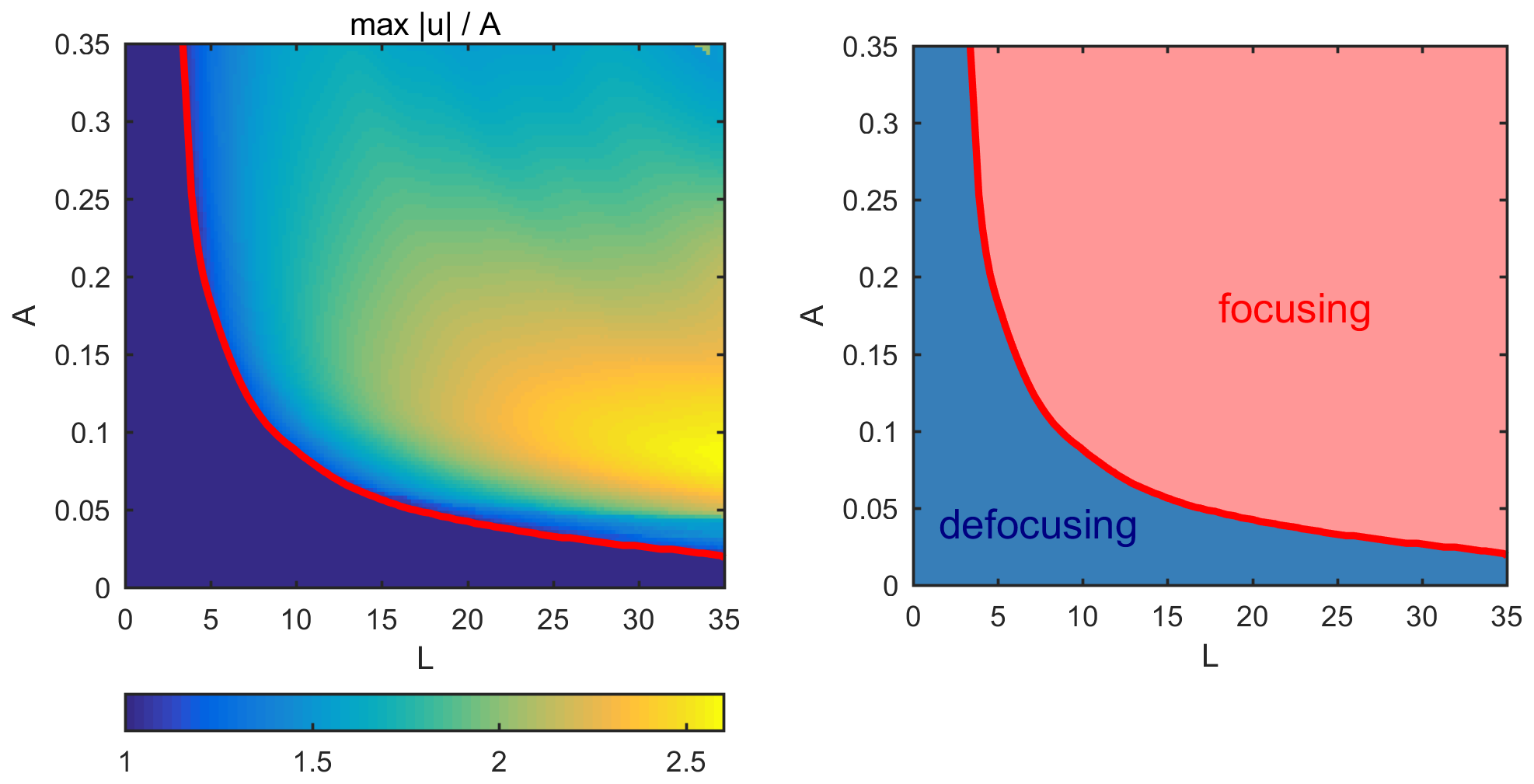} 
\caption{\textit{Left: }The wave group amplification factor $\dfrac{1}{A}
u_\text{max}(A,L)$ computed using  the reduced-order dynamical system for
the rare events  $u_r(x,t)$. \textit{Right:} Partition of $(A,L)$ plane into
focusing and defocusing regions according to the reduced order model.}
\label{fig:focusing}
\end{figure}

With this reduced order description of the dynamics for  the intermittent component $u_r$, we  investigate the evolution of various wavegroups   $(A,L)$ in order to quantify the conditions for the occurrence of rare transitions,   which will determine  $R_e$. More specifically, we consider initial wavegroups of the form
\begin{equation*}
u_{r}(x,t_c)=\Pi _{V_s} [A\sech( x/L ) ].
\end{equation*}
We   compute the value of the first spatiotemporal local maximum  of $u_r(x,t)$ for a range of $(A,L)$ to investigate group dynamics under different lengthscales  $L$ and amplitudes $A$. In the left pane of~\cref{fig:focusing} we display the value of this map  divided by the initial amplitude $A$. This wave group amplification factor is  $1$ for defocusing groups and greater than $1$ for groups that undergo nonlinear focusing. Importantly,  by this  map we can partition the space (right pane of~\cref{fig:focusing}) and subscribe the region where   $u_\text{max}(A,L)/A > 1$ as the critical conditions that cause  rare transitions, since for these states  $\lambda_T > 0$   and $u_r(x,t)$  (conditional on the background state $(A,L)$) is a focusing wave packet.  This gives the set 
\begin{equation} 
R_e = \biggl\{ (A,L)   \biggm\vert \frac{1}{A} u_\text{max}(A,L) >1 \biggr\},
\end{equation}
which   parameterizes the instability region in terms of just  two  parameters~$(A,L)$.

\subsection{Conditional statistics of rare events}
\label{sec:extremeResponse}

\begin{figure}[tb]
        \centering
        \includegraphics[width=0.8\textwidth]{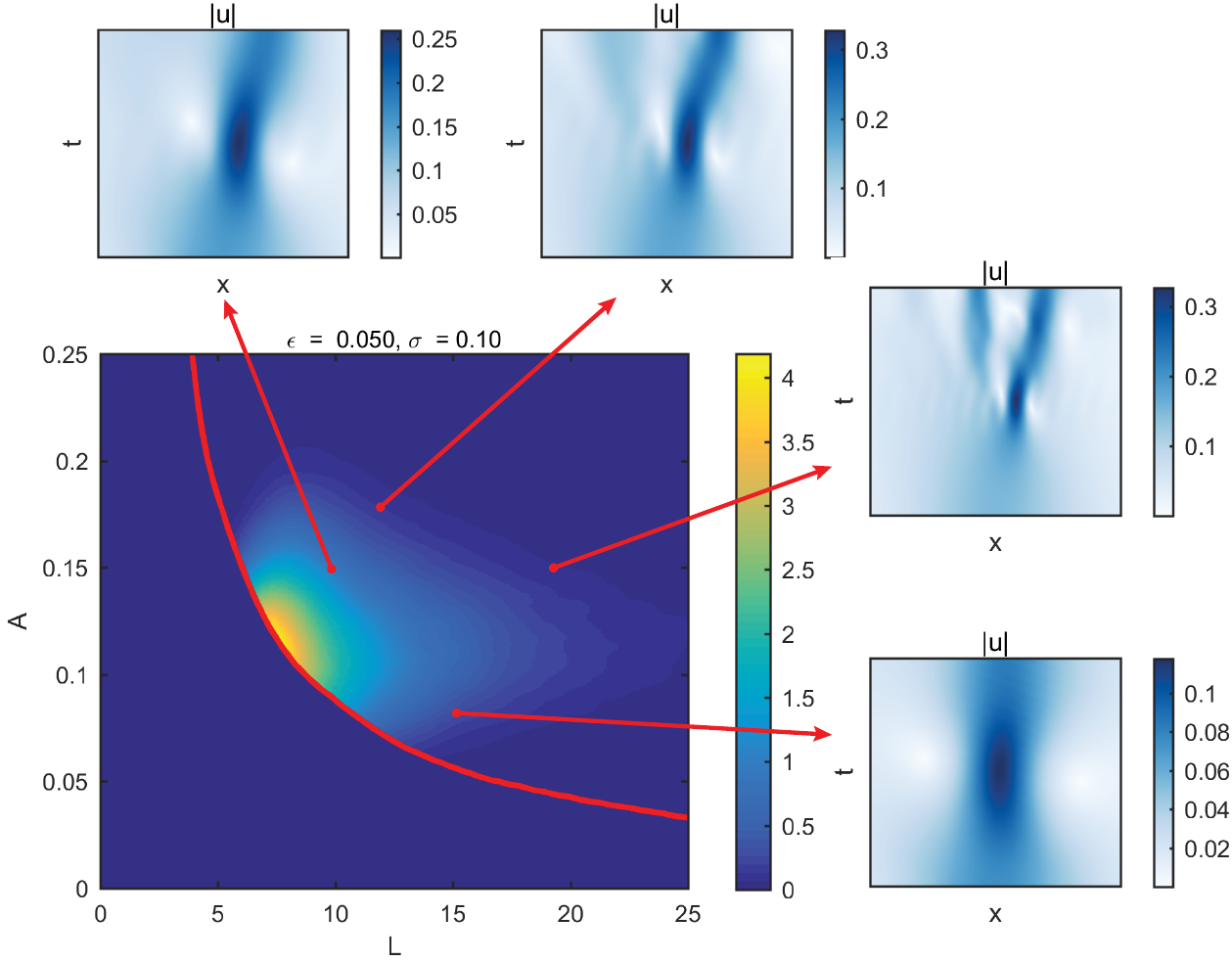}
        \caption{Procedure for computing the  conditional statistics for rare events.
 First we  determine the probability  of     wavegroups, $\pdf_{AL}(A,L)$, in $R_e$ for a given
spectrum. Then we  compute the pdf for the quantity of interest for each focusing group, using the reduced order model.}
        \label{fig:extGroupVis}
\end{figure}

To determine   $\pdf_{\abs{u}}(q \mid  \lVert u \rVert > \zeta,u_b \in R_e) $ we express it in terms of the wavegroup parameters, i.e. in the form    $\pdf_{\abs{u}}(q \mid\lVert u \rVert > \zeta , \, A, L\in R_e)$, where $A$ and $L$ denote the parameters  of  an arbitrary wavegroup formed in the stochastic background $u_b$ and lead to a rare event defined by $\zeta = \sigma_0$, the standard deviation of the initial spectrum. In the rare event regime we approximate $u\simeq u_r$. The estimation of this quantity involves simulations of $u_r$ through the reduced-order model. We perform the simulation until the wavegroup  relaxes back to the background state, necessarily a short time simulation. We also sample spatial points for which the response has important magnitude, in practice this means that we consider points $x \in [-2L,2L]$. This is done for all unstable wavegroup parameters $(A,L)$ for which the probability of occurrence $\pdf_{AL}(A,L)$ within the background dynamics is finite. A visual demonstration of this procedure is displayed in~\cref{fig:extGroupVis}.

We explicitly compute $\pdf_{AL}(A,L)$ by generating many random fields according
to~\eqref{eq:gaussSpec}. This random field represents typical realizations of the background state $u_b$, where the dominant effect is linear dispersion, which continuously mixes the phases between harmonics. We note that we ignore  the evolution of the spectrum due to weak nonlinearities (typically the spectrum tends to broaden~\cite{dysthe2003, xiao13})  and we use the initial spectrum as the  spectrum of the background $u_b$. This  simplification, which could be avoided by sampling the spectrum of the steady state using a relatively short simulation, does not cause any serious discrepancies to our final results. However, for more complex cases, such as two-dimensional nonlinear waves, where the spectrum is continuously evolving, the   transient character of the background statistics may have to be taken into consideration. Our framework can support such situation without any modifications. 

For each random field realization, we apply a group detection procedure (described in detail in~\cite{cousinsSapsis2015_JFM}), which returns a set of the groups in the field along with the amplitude and length scale of each group. This is a fast computation since it only requires generating and analyzing random realizations out of a given spectrum. After generating many random fields and computing the groups, we have a set of samples of $(A,L)$; subsequently we estimate the joint density $\pdf_{AL}(A,L)$, scaling the amplitudes in the resulting pdf by their corresponding lengthscale.  

\begin{figure}[tb]
        \centering
        \includegraphics[width=\textwidth]{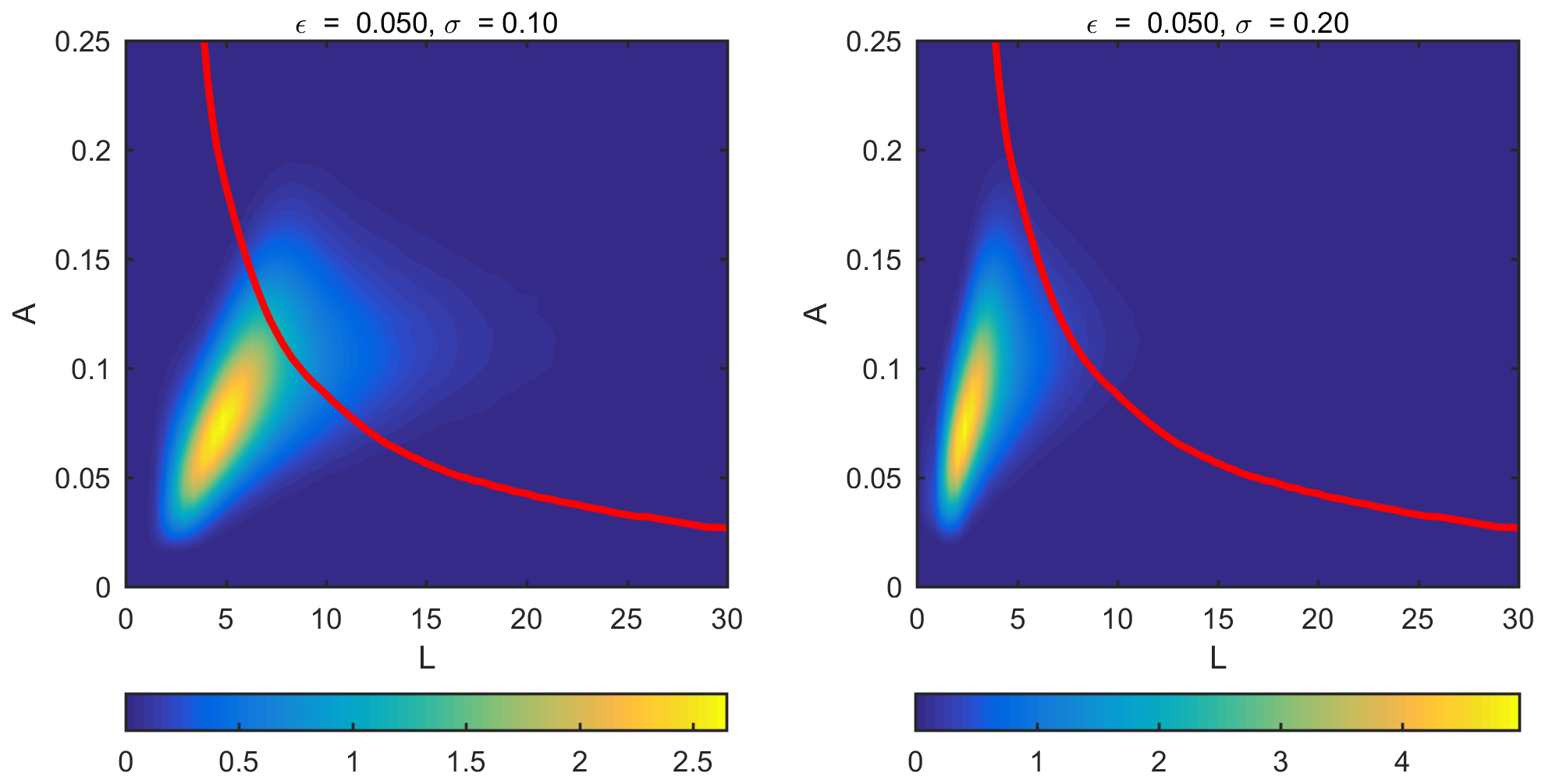}
        \caption{Group density $\pdf_{AL}(A,L)$ for Gaussian spectra with
$\epsilon=0.05$, $\sigma = 0.1$ (left) and $\epsilon=0.05$, $\sigma=0.2$
(right). The red curve is the $R_e$ boundary.} 
        \label{fig:groupDensity}
\end{figure}

Example group densities for two different spectra are displayed in~\cref{fig:groupDensity}.  In each figure we overlay the $R_e$ boundary.   Notice that in the case displayed on the right panel, the spectral width is larger, meaning that groups are narrower.  The effect of this change is to \emph{reduce} the number of focusing groups, and thus reduce the number of extreme waves.  This is consistent with the Benjamin-Feir Index ($2\sqrt{2} \epsilon/\sigma$), used in the nonlinear water-waves community to evaluate the  likelihood for  rare events,  which is reduced by increasing $\sigma$.

\subsection{Conditional statistics of background dynamics}\label{sec:mnls_regularcomponenet}

The next ingredient is to  estimate   the  conditional statistics for the background dynamics $\pdf_{\abs{u}}(q \mid u=u_{b})$. We will have only background components if and only if the occurring wavegroups belong in $R_e^c$. As a result, we have that  $\pdf_{\abs{u}}(q \mid u=u_{b})=\pdf_{\abs{u}}(q \mid A, L\in R_e^c).$ Moreover, in this regime we have predominantly linear dynamics and therefore the statistics of the background state can be approximated by a Gaussian distribution. More accurate representations that take into account weak nonlinearities~\cite{Tayfun1980} may also  be utilized to improve the accuracy of the main core of the distribution. 

For linear waves $\eta(t)$  with a  Gaussian distribution their envelope is Rayleigh distributed. Therefore, the conditional distribution for the quantity of interest in the background state is given by 
\begin{equation}\label{eq:stableregimedistribution}
 \pdf_{\abs{u}}(q\mid R_e^{\,c}) = \frac{q}{\sigma_s^2} e^{-q^2/ 2 \sigma_s^2},
\end{equation}
where $\sigma_s^2$ is the variance   of   \emph{non-focusing} wave groups, computed by taking  the variance of wave  groups in $R_e^{c}$. This is estimated   along side the computation described in~\cref{sec:mnls_likelihood} for determining $\pdf_{AL}(A,L)$, by   using the scale selection algorithm  to   identify the  wave  groups in $R_e^{c}$
 and then by taking the  variance of this set of wave packets.

\subsection{Probability for rare events}\label{sec:mnls_likelihood}

The final component  is the computation of the total probability to have a rare event in an arbitrary spatial location. We assume that the system is ergodic, and we use~\eqref{eq:rare_trans} in  the following spatial form 
\begin{equation} 
\mathbb{P}(\left\Vert u \right\Vert>\zeta,u_{b}\in
R_e)=\frac{1}{X_{D}}\int\limits_{u } \! X_{e}\left(u\right)\pdf_{u_b}(u)\,du
,
\end{equation}
where $X_{e}\left(u\right)$ is the spatial extend of each wavegroup associated with nonlinear focusing, and $X_D$ is the spatial size of the domain. Taking into account that the pdf $\pdf_{AL}$ already has been weighted with respect to the spatial extend of the wavegroups the last formula takes the form 
\begin{align}
        \mathbb{P}_{r}=\mathbb{P}(\left\Vert u \right\Vert>\zeta,u_{b}\in
R_e)   = \iint\limits_{R_e}\pdf_{AL}(A, L\mid R_e) \, dA dL.
        \label{eq:decompExtremeGroup}
\end{align}
This probability, of course, depends on the particular choice of the spectral properties of the field. Increasing  the spectral width, for example, will shift the distribution of $\rho_{AL}$ to the left, reducing in this way the total likelihood for the occurrence of a rare event (\cref{fig:groupDensity}).

\subsection{Synthesis and comparison to Monte-Carlo simulations}
\label{sec:envelopeSynthesis}

We have now determined all the components required to compute the heavy tailed    distribution for $\abs{u}$ according to  the decomposition-synthesis procedure. We have the following final result, by the total probability  probability argument:
\begin{equation}\label{eq:mnls_decompfinalresult}
        \pdf_{|u|}(q) = \pdf_{|u|}(q \mid R_e^{c}) ( 1-\mathbb{P}_r ) + \prob_{r}\iint\limits_{R_e}
\pdf_{|u|}(q \mid A,L) \pdf_{AL}(A,L)\, dAdL.
\end{equation}

\begin{figure}[htb]
        \centering
                \includegraphics{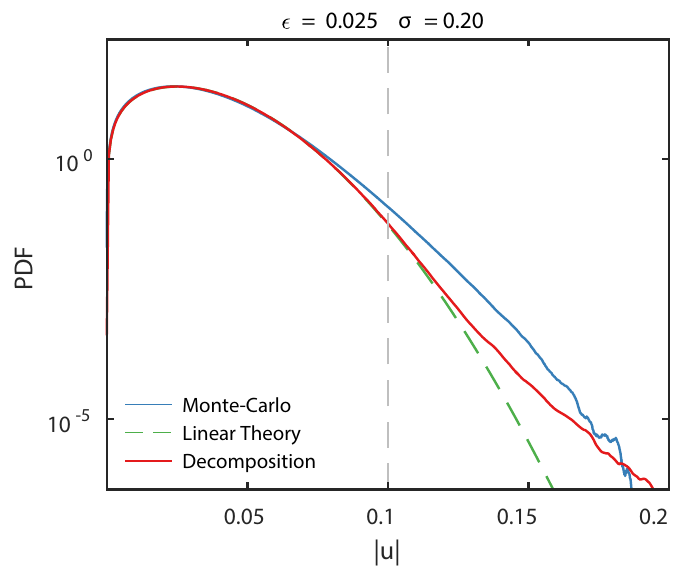}
                \includegraphics{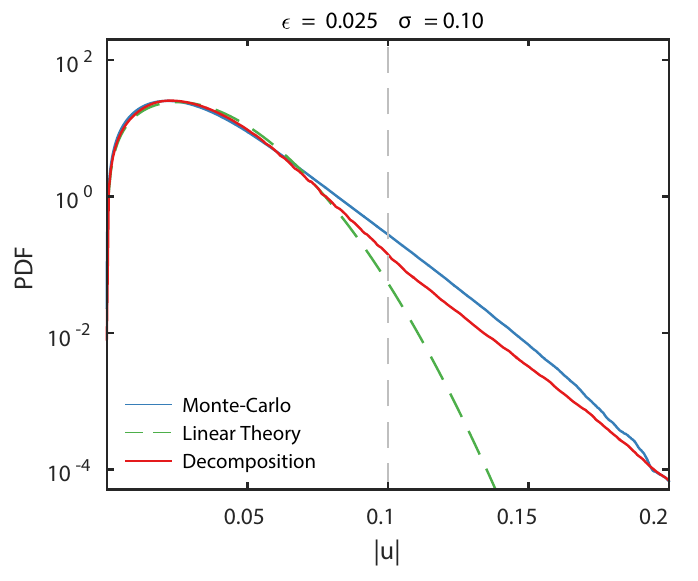}
                \includegraphics{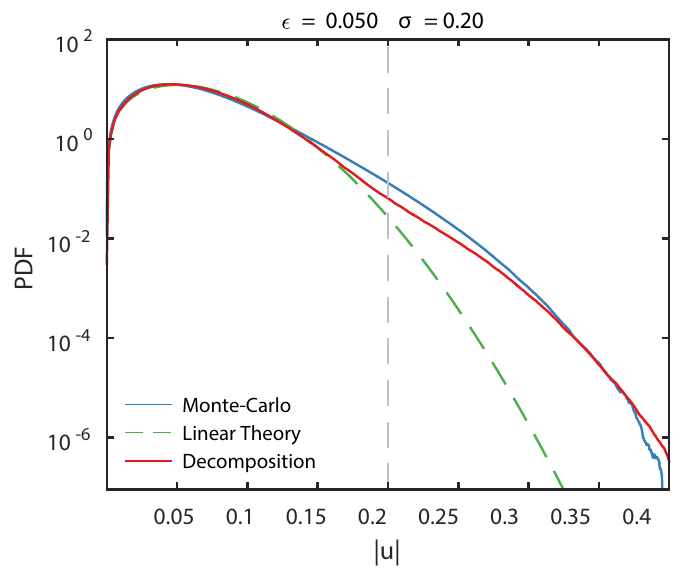}     
                \includegraphics{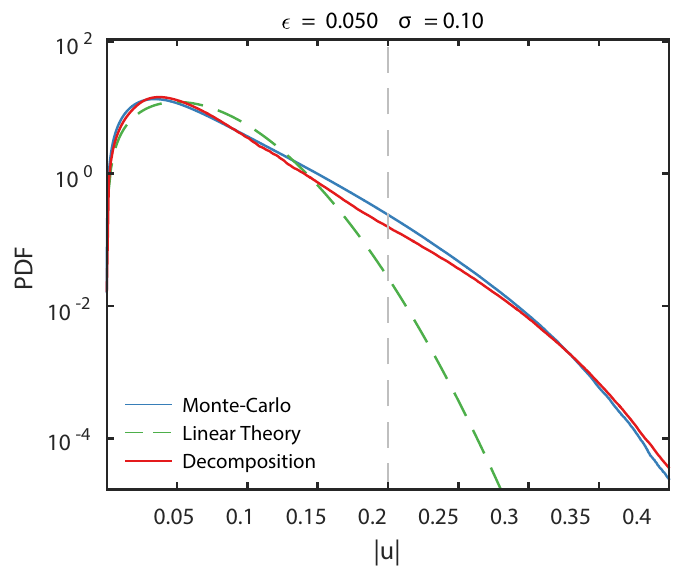}
        \caption{Probability distribution function approximation by the decomposition-synthesis method for the MNLS equation~\eqref{eq:MNLS}.  In each case, the blue curve denotes Monte Carlo simulation of the MNLS and the red line represents the approximation by the decomposition-synthesis procedure.  The vertical dashed line denotes $4\epsilon$ (i.e. 4 standard deviations) and $x$-axis plotted to $8\epsilon$.  We also display the Rayleigh distribution   of $|u|$  (green dashed line). Ordered in increasing BFI regimes: top left $\text{BFI} = 0.35 $, top right $\text{BFI} = 0.71$, bottom left $\text{BFI} = 0.71 $, bottom right $\text{BFI} = 1.41 $.} 
        \label{fig:exDecompEnv}
\end{figure}

In~\cref{fig:exDecompEnv} we demonstrate the decomposition-synthesis method on four different spectra alongside the `true' density  from Monte-Carlo simulations. To compute the Monte-Carlo statistics we use $10^{4}$ realizations of the MNLS equation on a spatial domain $256 \pi$ and a total sampled duration of $1000$ time units, discarding the first $500$ time units for cases with  $\sigma=0.10$ case and $500$ time units for the cases with $\sigma=0.20$ case. Overall, we see that the  decomposition-synthesis  approach gives good approximations to the true density.   For the most heavy-tailed cases, we observe that large waves occur many orders of magnitude more frequently than predicted by linear dynamics (Rayleigh distribution).  Our  method compares favorably with the Monte-Carlo results   at a fraction of the computational cost. 

We note that the computational savings of  the decomposition-synthesis  method is due to several important advantages over  the direct brute force Monte-Carlo method. First and very importantly, we do not have to wait for extreme waves to emerge from the background wave field; moreover, we do not have to wait until a sufficient number of extreme waves occur in order to obtain reliable tail statistics. This is because we simulate  extreme wave groups  directly by carefully choosing the initial/background conditions that trigger them. The fact that we use  a low dimensional  reduced order model to evolve  extreme wave groups, makes applying our estimation procedure very inexpensive  for computing  the  extreme components.  And finally, the distribution of the background dynamics was obtained analytically.  

We also emphasize that  once we performed the decomposition-synthesis procedure, computing the distribution for different background spectra is  extremely  cheap, since the only quantity that needs to be recomputed is the distribution of various groups $\pdf_{AL}(A,L)$ and $\sigma_s^2$, which is easy and fast to determine since it does not require the simulation of the original SPDE or the solution of the reduced-order model.

\section{Conclusions}\label{sec:conc}
We have considered the problem of  quantifying rare event statistics due to internal and finite-time instabilities in general dynamical systems. Our analysis is based on the assumption that the conditional dynamics involving rare events are of low-dimensionality, although the total response of the system can still be very high dimensional.  Relying on this setup, we have formulated a conditional decomposition into low-dimensional extreme events caused by internal instabilities, and a high dimensional stochastic background. This decomposition allows for the study of the two components separately (but taking into account mutual interactions), using \emph{different} uncertainty quantification methods that (i)  take into consideration the possibly high-dimensional (broad spectrum) character of the stochastic background, and (ii) the nonlinear and unstable character of the rare events. 

The adopted decomposition is in full correspondence with a partition of the phase space into a stable region, where we have no internal instabilities, and a region where non-linear instabilities  lead to extreme transitions with high probability. We quantify the statistics in the stable region using a Gaussian approximation, while the non-Gaussian distributions associated with the intermittently unstable regions of   the phase space, are inexpensively computed through order-reduction methods that take into account the strongly nonlinear character of the dynamics. The probabilistic information for the two domains is analytically synthesized through a total probability argument.

The proposed approach allows for the derivation of the statistics for any quantity of interest in a semi-analytical form, where only a few carefully selected simulations through a reduced order model are sufficient for the accurate determination of the heavy tail structure. For low-dimensional systems the developed framework allows for the derivation of fully analytical forms, while for more complex systems it provides an inexpensive computational method to determine extreme event statistics.   

To demonstrate the new method we considered two systems of increasing complexity where non-trivial energy exchanges occur due to internal instabilities, leading to extreme responses. The first application was a two-degree-of-freedom nonlinear system of coupled mechanical oscillators encountered in a viariety of engineering settings. This setup leads to non-Gaussian statistics with heavy tails characterized by qualitatively different regimes. Through numerical experiments we demonstrated  that our method is able to capture   very accurately the core of the response distribution, the  exponential like heavy-tails at extreme values,  and the subsequent sub-exponential decay at very extreme values of distribution.

The second application was a prototype nonlinear envelope equation, that describes the  one-dimensional propagation of deep water waves, where   extreme waves (known as rogue or freak waves)  randomly appear due to  nonlinear focusing in a wavefield. This is an example that is particularly challenging due to its nonlinear, dispersive, and infinite dimensional character. In such case our method is very advantageous as it allows  to   separately quantify the extreme wavegroups from  the background field.      Comparisons with direct   Monte-Carlo simulation  demonstrated the effectiveness of our approach on semi-analytically and inexpensively capturing the heavy-tailed statistics   for   the distribution of   the local wave field maxima, for four different spectra of increasing  heavy-tailed statistics . We also demonstrated the value of our approach on computing the pdf of interest  for different spectrum parameters  in comparison with Monte-Carlo results where the simulation would have to be run anew.
 
\section*{Acknowledgments}
TPS has been supported through the Office of Naval Research grant ONR N00014-14-1-0520, the Army Research Office Young Investigator Award 66710-EG-YIP, and the DARPA grant HR0011-14-1-0060. MAM and WC have been partially supported  by the first and second grants. 

\bibliographystyle{plain}
\bibliography{paper}

\begin{thebibliography}{10}

\bibitem{arnold_l}
L.~Arnold, I.~Chueshov, and G.~Ochs.
\newblock {Stability and capsizing of ships in random sea - A survey}.
\newblock {\em Nonlinear Dynamics}, 36:135--179, 2004.

\bibitem{belenky07}
V.~L. Belencky and N.~B. Sevastianov.
\newblock {\em {Stability and Safety of Ships: Risk of Capsizing}}.
\newblock The Society of Naval Architects and Marine Engineers, 2007.

\bibitem{Bolhuis2003}
P.~G. Bolhuis.
\newblock {Transition-path sampling of beta-hairpin folding.}
\newblock {\em Proceedings of the National Academy of Sciences of the United
  States of America}, 100(21):12129--34, oct 2003.

\bibitem{caughey82}
T.~K. Caughey and F.~Ma.
\newblock {The exact steady-state solution of a class of non-linear stochastic
  systems}.
\newblock {\em Int. J. Non. Linear. Mech.}, 17:137--142, 1982.

\bibitem{chow92}
P.-L. Chow.
\newblock {Some parabolic Ito equations}.
\newblock {\em Comm. Pure Appl. Math.}, 45:97, 1992.

\bibitem{cousins_sapsis}
W.~Cousins and T.~P. Sapsis.
\newblock {Quantification and prediction of extreme events in a one-dimensional
  nonlinear dispersive wave model}.
\newblock {\em Physica D}, 280:48--58, 2014.

\bibitem{cousinsSapsis2015_JFM}
W.~Cousins and T.~P. Sapsis.
\newblock {Reduced order prediction of rare events in unidirectional nonlinear
  water waves}.
\newblock {\em Submitted to Journal of Fluid Mechanics}, 2015.

\bibitem{cousinsSapsis2015_PRE}
W.~Cousins and T.~P. Sapsis.
\newblock {The unsteady evolution of localized unidirectional deep water wave
  groups}.
\newblock {\em Physical Review E}, 91:063204, 2015.

\bibitem{dembo00}
A.~Dembo and O.~Zeitouni.
\newblock {\em {Large Deviations Techniques and Applications}}.
\newblock Springer-Verlag, New York, 2000.

\bibitem{Dysthe08}
K.~Dysthe, H.~E. Krogstad, and P.~M{\"{u}}ller.
\newblock {Oceanic Rogue Waves}.
\newblock {\em Annu. Rev. Fluid Mech.}, 40(1):287, 2008.

\bibitem{dysthe1979}
K.~B. Dysthe.
\newblock {Note on a modification to the nonlinear Schrodinger equation for
  application to deep water waves}.
\newblock {\em Proceedings of the Royal Society of London. A. Mathematical and
  Physical Sciences}, 369(1736):105--114, 1979.

\bibitem{dysthe2003}
K.~B. Dysthe, K.~Trulsen, H.~E. Krogstad, and H.~Socquet-Juglard.
\newblock {Evolution of a narrow-band spectrum of random surface gravity
  waves}.
\newblock {\em J. Fluid Mech}, 478:1--10, 2003.

\bibitem{E.2006}
W.~E. and E.~Vanden-Eijnden.
\newblock {Towards a Theory of Transition Paths}.
\newblock {\em Journal of Statistical Physics}, 123(3):503--523, may 2006.

\bibitem{E2010}
W.~E and E.~Vanden-Eijnden.
\newblock {Transition-path theory and path-finding algorithms for the study of
  rare events.}
\newblock {\em Annual review of physical chemistry}, 61:391--420, jan 2010.

\bibitem{Easterling2000}
D.~R. Easterling.
\newblock {Climate Extremes: Observations, Modeling, and Impacts}.
\newblock {\em Science}, 289(5487):2068--2074, sep 2000.

\bibitem{Embrechts12}
P.~Embrechts, C.~Kluppelberg, and T.~Mikosch.
\newblock {\em {Modeling Extremal Events}}.
\newblock Springer, 2012.

\bibitem{epstein69}
E.~S. Epstein.
\newblock {Stochastic Dynamic Predictions}.
\newblock {\em Tellus}, 21:739--759, 1969.

\bibitem{Eyring1935}
H.~Eyring.
\newblock {The Activated Complex and the Absolute Rate of Chemical Reactions.}
\newblock {\em Chemical Reviews}, 17(1):65--77, aug 1935.

\bibitem{Frei2001}
C.~Frei and C.~Sch{\"{a}}r.
\newblock {Detection Probability of Trends in Rare Events: Theory and
  Application to Heavy Precipitation in the Alpine Region}.
\newblock {\em Journal of Climate}, 14(7):1568--1584, apr 2001.

\bibitem{freidlin98}
M.~I. Freidlin and A.~D. Wentzell.
\newblock {\em {Random Perturbations of Dynamical Systems}}.
\newblock Berlin Springer Verlag, 2nd editio edition, 1998.

\bibitem{Galambos}
J.~Galambos.
\newblock {\em {The asymptotic theory of extreme order statistics}}.
\newblock Wiley Series in Probability and Statistics, 1978.

\bibitem{hense06}
A.~Hense and P.~Friederichs.
\newblock {Wind and Precipitation Extremes in the Earth’s Atmosphere}.
\newblock In {\em Extreme Events in Nature and Society}, pages 169--187.
  Springer-Verlag, 2006.

\bibitem{Holmes_et_al96}
P.~Holmes, J.~Lumley, and G.~Berkooz.
\newblock {\em {Turbulence, Coherent Structures, Dynamical Systems and
  Symmetry}}.
\newblock Cambridge University Press, 1996.

\bibitem{Kharif2003}
C.~Kharif and E.~Pelinovsky.
\newblock {Physical mechanisms of the rogue wave phenomenon}.
\newblock {\em European Journal of Mechanics, B/Fluids}, 22(6):603--634, 2003.

\bibitem{Kishore12}
V.~Kishore, M.~S. Santhanam, and R.~E. Amritkar.
\newblock {Extreme events and event size fluctuations in biased random walks on
  networks}.
\newblock {\em Physical Review E}, 85:56120, 2012.

\bibitem{Kreuzer}
E.~Kreuzer and W.~Sichermann.
\newblock {The effect of sea irregularities on ship rolling}.
\newblock {\em Computing in Science and Engineering}, May/June:26--34, 2006.

\bibitem{kundur94}
P.~Kundur.
\newblock {\em {Power System Stability and Control}}.
\newblock McGraw-Hill Education, 1994.

\bibitem{Leadbetter}
M.~R. Leadbetter, G.~Lindgren, and H.~Rootzen.
\newblock {\em {Extremes and related properties of random sequences and
  processes}}.
\newblock Springer, New York, 1983.

\bibitem{majda_branicki_DCDS}
A.~J. Majda and M.~Branicki.
\newblock {Lessons in Uncertainty Quantification for Turbulent Dynamical
  Systems}.
\newblock {\em Discrete and Continuous Dynamical Systems}, 32:3133--3221, 2012.

\bibitem{Majda_filter}
A.~J. Majda and J.~Harlim.
\newblock {\em {Filtering Complex Turbulent Systems}}.
\newblock Cambridge University Press, 2012.

\bibitem{Majda2014a}
A.~J. Majda and Y.~Lee.
\newblock {Conceptual dynamical models for turbulence.}
\newblock {\em Proceedings of the National Academy of Sciences of the United
  States of America}, 111(18):6548--53, may 2014.

\bibitem{majda1997}
A.~J. Majda, D.~W. McLaughlin, and E.~G. Tabak.
\newblock {A one-dimensional model for dispersive wave turbulence}.
\newblock {\em Journal of Nonlinear Science}, 7(1):9--44, 1997.

\bibitem{Majda2014}
A.~J. Majda, D.~Qi, and T.~P. Sapsis.
\newblock {Blended particle filters for large-dimensional chaotic dynamical
  systems}.
\newblock {\em Proceedings of the National Academy of Sciences},
  111(21):7511--7516, 2014.

\bibitem{Metzner2009}
P.~Metzner, C.~Sch{\"{u}}tte, and E.~Vanden-Eijnden.
\newblock {Transition Path Theory for Markov Jump Processes}.
\newblock {\em Multiscale Modeling {\&} Simulation}, 7(3):1192--1219, jan 2009.

\bibitem{Metzner2006}
Philipp Metzner, Christof Sch{\"{u}}tte, and Eric Vanden-Eijnden.
\newblock {Illustration of transition path theory on a collection of simple
  examples.}
\newblock {\em The Journal of chemical physics}, 125(8):084110, aug 2006.

\bibitem{mohamad2015}
M.~A. Mohamad and T.~P. Sapsis.
\newblock {Probabilistic description of extreme events in intermittently
  unstable systems excited by correlated stochastic processes}.
\newblock {\em SIAM ASA J. of Uncertainty Quantification}, 3:709--736, 2015.

\bibitem{Naess1982}
A~Naess.
\newblock {Extreme value estimates based on the envelope concept}.
\newblock {\em Applied Ocean Research}, 4(3):181, 1982.

\bibitem{naess_book}
A.~Naess and T.~Moan.
\newblock {\em {Stochastic Dynamics of Marine Structures}}.
\newblock Cambridge University Press, 2012.

\bibitem{extreme_value09}
M.~Nicodemi.
\newblock {Extreme value statistics}.
\newblock {\em in Encyclopedia of complexity and systems science}, E:3317,
  2009.

\bibitem{Olagnon2005}
M.~Olagnon and M.~Prevosto.
\newblock {\em {Rogue Waves 2004: Proceedings of a Workshop Organized by
  Ifremer and Held in Brest, France, 20-21-22 October 2004, Within the Brest
  Sea Tech Week 2004}}.
\newblock Editions Quae, 2005.

\bibitem{paik03}
J.~K. Paik and A.~K. Thayamballi.
\newblock {\em {Ultimate Limit State Design of Steel-Plated Structures}}.
\newblock John Wiley and Sons, 2003.

\bibitem{Pourbeik}
P.~Pourbeik, P.~Kundur, and C.~Taylor.
\newblock {The anatomy of a power grid blackout - Root causes and dynamics of
  recent major blackouts,}.
\newblock {\em IEEE Power and Energy Magazine}, 4(5):22--29, 2006.

\bibitem{Pratt1986}
L.~R. Pratt.
\newblock {A statistical method for identifying transition states in high
  dimensional problems}.
\newblock {\em The Journal of Chemical Physics}, 85(9):5045, nov 1986.

\bibitem{qi15}
D.~Qi and A.~J. Majda.
\newblock {Predicting Fat-Tailed Intermittent Probability Distributions in
  Passive Scalar Turbulence with Imperfect Models through Empirical Information
  Theory}.
\newblock {\em Submitted to Physica D}, 2015.

\bibitem{Thomas_extremes}
R.-D. Reiss and M.~Thomas.
\newblock {\em {Statistical analysis of extreme values}}.
\newblock Birkh{\{}{\"{a}}{\}}user; 3rd edition, 2007.

\bibitem{sapsis11a}
T.~P. Sapsis.
\newblock {Attractor local dimensionality, nonlinear energy transfers, and
  finite-time instabilities in unstable dynamical systems with applications to
  2D fluid flows}.
\newblock {\em Proceedings of the Royal Society A}, 469(2153):20120550, 2013.

\bibitem{SapsisLermusiaux09}
T.~P. Sapsis and P.~F.~J. Lermusiaux.
\newblock {Dynamically Orthogonal field equations for continuous stochastic
  dynamical systems}.
\newblock {\em Physica D}, 238:2347--2360, 2009.

\bibitem{sapsis_majda_mqg}
T.~P. Sapsis and A.~J. Majda.
\newblock {A statistically accurate modified quasilinear Gaussian closure for
  uncertainty quantification in turbulent dynamical systems}.
\newblock {\em Physica D}, 252:34--45, 2013.

\bibitem{sapsis_majda_qgdo}
T.~P. Sapsis and A.~J. Majda.
\newblock {Blended reduced subspace algorithms for uncertainty quantification
  of quadratic systems with a stable mean state}.
\newblock {\em Physica D}, 258:61, 2013.

\bibitem{sapsis_majda_mqgdo}
T.~P. Sapsis and A.~J. Majda.
\newblock {Blending Modified Gaussian Closure and Non-Gaussian Reduced Subspace
  methods for Turbulent Dynamical Systems}.
\newblock {\em Journal of Nonlinear Science}, 23:1039, 2013.

\bibitem{sapsis_majda_tur}
T.~P. Sapsis and A.~J. Majda.
\newblock {Statistically Accurate Low Order Models for Uncertainty
  Quantification in Turbulent}.
\newblock {\em Proceedings of the National Academy of Sciences},
  110:13705--13710, 2013.

\bibitem{Shadman}
D.~Shadman and B.~Mehri.
\newblock {A non-homogeneous Hill's equation}.
\newblock {\em Applied Mathematics and Computation}, 167:68--75, 2005.

\bibitem{Sirovich87}
L.~Sirovich.
\newblock {Turbulence and the dynamics of coherent structures, Parts I, II and
  III}.
\newblock {\em Quart. Appl. Math.}, XLV:561--590, 1987.

\bibitem{Sobczyk91}
K.~Sobczyk.
\newblock {\em {Stochastic Differential Equations}}.
\newblock Kluwer Academic Publishers, Dordrecht, The Netherlands, 1991.

\bibitem{soize88}
C.~Soize.
\newblock {Steady-state solution of Fokker-Planck equation in higher
  dimension}.
\newblock {\em Probab. Eng. Mech.}, 3:196--206, 1988.

\bibitem{soize94}
C.~Soize.
\newblock {\em {The Fokker-Planck equation for stochastic dynamical systems and
  its explicit steady state solutions}}.
\newblock World {\{}S{\}}cientific, 1994.

\bibitem{Soong_Grigoriou93}
T.~Soong and M.~Grigoriu.
\newblock {\em {Random Vibration of Mechanical and Structural Systems}}.
\newblock PTR Prentice Hall, 1993.

\bibitem{sowers92}
R.~Sowers.
\newblock {Large Deviations for a reaction diffusion equation with non-Gaussian
  perturbations}.
\newblock {\em Ann. Probab.}, 20:504, 1992.

\bibitem{sritharan06}
S.~S. Sritharan and P.~Sundar.
\newblock {Large deviations for the two-dimensional Navier–Stokes equations
  with multiplicative noise}.
\newblock {\em Stochastic Processes and their Applications}, 116:1636, 2006.

\bibitem{stroock84}
D.~Stroock.
\newblock {\em {An Introduction to the Theory of Large Deviations}}.
\newblock Springer-Verlag, New York, 1984.

\bibitem{Susuki2012a}
Y.~Susuki and I.~Mezic.
\newblock {Nonlinear Koopman Modes and a Precursor to Power System Swing
  Instabilities}.
\newblock {\em IEEE Transactions on Power Systems}, 27(3):1182--1191, aug 2012.

\bibitem{Tayfun1980}
M.~A. Tayfun.
\newblock {Narrow-band nonlinear sea waves}.
\newblock {\em Journal of Geophysical Research}, 85(C3):1548, 1980.

\bibitem{tong15}
X.~Tong and A.~J. Majda.
\newblock {Intermittency in Turbulent Diffusion Models with a Mean Gradient}.
\newblock {\em Submitted to Nonlinearity}, 2015.

\bibitem{trulsen1996}
K.~Trulsen and K.~B. Dysthe.
\newblock {A modified nonlinear Schr{\"{o}}dinger equation for broader
  bandwidth gravity waves on deep water}.
\newblock {\em Wave motion}, 24(3):281--289, 1996.

\bibitem{varadhan84}
S.~R.~S. Varadhan.
\newblock {\em {Large Deviations and Applications}}.
\newblock SIAM, 1984.

\bibitem{varadhan08}
S.~R.~S. Varadhan.
\newblock {Special invited paper: Large deviations}.
\newblock {\em The Annals of Probability}, 36:397, 2008.

\bibitem{wang00}
R.~Wang and Z.~Zhang.
\newblock {Exact stationary solutions of the Fokker-Planck equation for
  nonlinear oscillators under stochastic parametric and external excitations}.
\newblock {\em Nonlinearity}, 13:907--920, 2000.

\bibitem{Wigner1938}
E.~Wigner.
\newblock {The transition state method}.
\newblock {\em Transactions of the Faraday Society}, 34:29, jan 1938.

\bibitem{xiao13}
W.~Xiao, Y.~Liu, G.~Wu, and D.~K.~P. Yue.
\newblock {Rogue wave occurrence and dynamics by direct simulations of
  nonlinear wave-field evolution}.
\newblock {\em Journal of Fluid Mechanics}, 720:357--392, 2013.

\bibitem{zhu90}
W.~Q. Zhu.
\newblock {Exact Solutions for Stationary Responses of Several Classes of
  Nonlinear Systems to Parametric and/or External White Noise Excitations}.
\newblock {\em Appl. Mathematics Mech. Ed.}, 11:165--175, 1990.

\end{thebibliography}

\end{document}